\documentclass[12pt]{article}

\usepackage{latexsym}  
\def\a{\alpha} 
\def\b{\beta}  
\def\g{\gamma}
\def\G{\Gamma}   
   
\def\d{\delta}

\def\l{\lambda}
\def\m{\mu} 
\def\n{\nu}   
\def\h{\eta}
\def\f{\phi}   
\def\r{\rho}   
\def\o{\omega}  
\def\s{\sigma}

\def\x{\xi}   
\def\e{\varepsilon}

\def\pa{\partial} 
  
\def\ll{\left}  
\def\rr{\right}  
\def\be{\begin{equation}}   
\def\ee{\end{equation}}   
\def\bqn{\begin{eqnarray}}   
\def\eqn{\end{eqnarray}}
\def\nn{\nonumber}

\def\cc{{\cal C}}  
\def\cd{{\cal D}}

\def\ci{{\cal I}}  
  
\def\ck{{\cal K}}  
\def\cl{{\cal L}}

\def\co{{\cal O}}

\def\car{{\cal R}}

\def\cw{{\cal W}}

\newtheorem{theorem}{Theorem}[subsubsection]  
\newtheorem{lemma}{Lemma}[subsubsection]

\begin{document} 
\newcommand{\volume}{X}              %sets current volume,
\newcommand{\xyear}{2001}            %sets year in header
\newcommand{\issue}{X}               %sets current issue,
\newcommand{\recdate}{dd.mm.yyyy}    %sets received date,
\newcommand{\revdate}{dd.mm.yyyy}    %sets revised date,
\newcommand{\revnum}{0}              %number of revisions,
\newcommand{\accdate}{dd.mm.yyyy}    %sets accepted date,
\newcommand{\coeditor}{ue}           %sets (co)editor,
\newcommand{\firstpage}{1}           %first page number,
\newcommand{\lastpage}{31}            %last page number,
\setcounter{page}{\firstpage}        %sets page counter to first page number
\newcommand{\keywords}{Consistent interactions, Weyl gravity, BRST cohomology}
%%%%%%%%%%%%%%%%%%%%%%%%%%%%%%%%%%%%%%%%%%%%%%%%%%%%%%%%%%%%%%%%%%%%%%%%%%%%
%%%%%%%%%%%%%%%% please give up to three PACS numbers here %%%%%%%%%%%%%%%%%
%%%%%%%%%%%%%%%%%%%%%%%%%%%%%%%%%%%%%%%%%%%%%%%%%%%%%%%%%%%%%%%%%%%%%%%%%%%%
%\newcommand{\PACS}{04.20.-q,95.30.Sf}
%\newcommand{\PACS}{01.30.Rr, 42.15.Fr, 96.60.Bn}%%%%%%%%%%%%%%%%%%%%%%%%%%%
%%%%%%%%%%%% must not exceed 80 characters in length together %%%%%%%%%%%%%%
%\newcommand{\shorttitle}{N.\ Boulanger and M.\ Henneaux, 
%A derivation of Weyl gravity }

%\title{A derivation of Weyl gravity}  
%\author{Nicolas Boulanger$^{1,}$\footnote{``Chercheur F.R.I.A.", Belgium},  
%and Marc Henneaux$^{1,2}$}  

%\newcommand{\address}
%  {$^{1}$Physique Th\'eorique et Math\'ematique,  Universit\'e Libre  
%   de Bruxelles,  C.P. 231, B-1050, Bruxelles, 
%   \\ \hspace*{0.5mm} Belgium      \\   
%    $^{2}$ Centro de Estudios Cient\'{\i}ficos, Casilla 1469, Valdivia, 
%    \\ \hspace*{0.5mm} Chile} 

%\newcommand{\email}{\tt nboulang@ulb.ac.be}   
%\maketitle
\begin{titlepage} 
\begin{flushright}
ULB-TH/01-15\\
\end{flushright}
\vskip .5cm
\begin{centering}

{\huge {\bf A derivation of Weyl gravity}}

\vspace{1cm}

{\Large Nicolas Boulanger$^{a,}$\footnote{``Chercheur F.R.I.A.", Belgium},
and Marc Henneaux$^{a,b}$} \\
\vspace{.4cm}
$^a$ Physique Th\'eorique et Math\'ematique,  Universit\'e Libre
de Bruxelles,  C.P. 231, B-1050, Bruxelles, Belgium      \\
\vspace{.2cm}
$^b$ Centro de Estudios Cient\'{\i}ficos, Casilla 1469, Valdivia, Chile
\vspace{.5cm} \\                                      

{\tt nboulang@ulb.ac.be,henneaux@ulb.ac.be}

\vspace{1cm}

\end{centering}      
\begin{abstract}  
In this paper, two things are done.
(i) Using cohomological techniques, we explore the consistent deformations of 
linearized conformal gravity in 4 dimensions. We show that the only possibility
involving no more than $4$ derivatives of the metric (i.e., terms
of the form $\partial^4 g_{\mu \nu}$, $\partial^3 g_{\mu \nu}
\partial g_{\alpha \beta}$, $\partial^2 g_{\mu \nu} 
\partial^2g_{\alpha \beta}$,
$\partial^2 g_{\mu \nu} \partial g_{\alpha \beta} \partial g_{\rho \sigma}$
or $\partial g_{\mu \nu} \partial g_{\alpha \beta} \partial g_{\rho \sigma}
\partial g_{\gamma \delta}$ with coefficients that involve undifferentiated
metric components - or terms with less derivatives) 
is given by the Weyl action 
$\int d^4x \sqrt{-g} W_{\a\b\g\d} W^{\a\b\g\d}$,
in much the same way as the Einstein-Hilbert action describes the only 
consistent manner to make a Pauli-Fierz massless spin-2 field self-interact
with no more than $2$ derivatives.  No a priori requirement of invariance
under diffeomorphisms is imposed: this follows automatically from
consistency.
(ii) We then turn to ``multi-Weyl graviton" theories. We show the impossibility
to introduce cross-interactions between the different types of Weyl
gravitons if one requests that the action reduces, in the free limit, to a
sum of linearized Weyl actions.  However, if different free limits are
authorized, cross-couplings become possible. An explicit example is given.
We discuss also how the results extend to other spacetime dimensions.
\end{abstract}  

\vfill
\end{titlepage}

\section{Introduction}  
\setcounter{equation}{0}  
\setcounter{theorem}{0}  
\setcounter{lemma}{0}  

\subsection{From the linearized Weyl action to the full Weyl action}

The discovery by Becchi, Rouet and Stora \cite{BRS} and Tyutin 
\cite{Tyutin} of the symmetry that now bears their names (``BRST
symmetry") is a landmark in the development of local gauge field
theories since it has brought in powerful algebraic methods
which have shed a new and deeper light on the structures
underlying renormalization and anomalies.  The application of the BRST
approach is, however, not confined to the quantum domain, since
it is quite useful already classically.  For instance, it can 
be used to analyse higher-order conservation laws
for Yang-Mills gauge models and Einstein gravity 
\cite{Barnich:1995db,Barnich:1995mt,BBHEinst}.  
It is also relevant to the problem of consistent
interactions, where it provides a cohomological reformulation
of the Noether method \cite{Barn-Henn}.  In this paper, 
we analyse the problem of
consistent deformations of linearized Weyl gravity from
the BRST standpoint.

There exist many ways to arrive at the Einstein equations
\cite{MTW}. One of them 
starts with the free action for a massless spin-2 field $h^{PF}_{\m\n}$ 
(Pauli-Fierz action \cite{Fierz:1939ix})
\bqn
S^{PF}_0[h^{PF}_{\mu \nu}] &=&  \int d^4 x \ll[  
-\frac{1}{2}\ll(\pa_{\m}{h}^{PF}_{\n\r}\rr)\ll(\pa^{\m}{h}^{PF\n\r}\rr)  
+\ll(\pa_{\m}{h}^{PF\m}_{~\n}\rr)\ll(\pa_{\r}{h}^{PF\r\n}\rr)\rr.\nn\\  
&&\ll.-\ll(\pa_{\n}{h}^{PF\m}_{~\m}\rr)\ll(\pa_{\r}{h}^{PF\r\n}\rr)  
+\frac{1}{2}\ll(\pa_{\m}{h}^{PF\n}_{~\n}\rr)  
\ll(\pa^{\m}{h}^{PF\r}_{~\r}\rr)\rr] \, . 
\label{PF}
\eqn
and investigates the consistent manners to self-couple $h^{PF}_{\m\n}$.
Under reasonable additional conditions, one can show that the only 
possibility is described by the Einstein-Hilbert action 
\be
S_0^{PF}[h^{PF}_{\m\n}] \rightarrow S^{EH}[g_{\m\n}]=\frac{2}{\kappa^2}
\int d^4x \sqrt{-g}(R-2\Lambda)\, ,~~~~g_{\m\n}=\h_{\m\n}+
\kappa{h^{PF}_{\m\n}}
\label{EH}
\ee
where $R$ is the scalar curvature of the metric $g_{\mu \nu}$ and
$g$ its determinant.  In (\ref{EH}), the coupling constant
$\kappa$ is of mass dimension $-1$.
At the same time, the abelian gauge invariance of (\ref{EH}), namely,
\be
\d_{\x}h^{PF}_{\m\n}=\pa_{\m}\x_{\n}+\pa_{\n}\x_{\m}
\ee
becomes elevated to diffeomorphism invariance,
\be
\frac{1}{\kappa} \d_{\x}g_{\m\n}=\x_{\m;\n}+\x_{\n;\m}.
\ee
This approach to the Einstein theory, which has a long history 
\cite{all,BDGH}, exhibits clearly the deep connection between
 massless spin-2 fields and diffeomorphism invariance.

In connection with the AdS/CFT correspondance, there has been renewed interest
recently in Weyl gravity \cite{0007211}, described by the conformally 
invariant action
\be
S= \frac{1}{2\a^2}\int d^4x \sqrt{-g} W_{\a\b\g\d} W^{\a\b\g\d} ~,
\label{actionconf}
\ee
where $ W^{\a}_{~\b\g\d}$ is the conformally invariant Weyl tensor,
\be\label{linearizedWeyl}
W^{\a}_{~\b\g\d}=R^{\a}_{~\b\g\d}-2\left(
\d^{\a}_{~[\g}K_{\d]\b}-g_{\b[\g}K_{\d]}^{~\a} \right)
\label{confweyltens}
\ee
(for an other recent work on Weyl gravity, see also \cite{Schmidt}).

We assign weight 1 to symmetrized and antisymmetrized
expressions.
Here, $K_{\a\b}$ is defined through
$K_{\a\b}=\frac{1}{n-2}\left(R_{\a\b}-\frac{1}{2(n-1)}g_{\a\b}
R \right)$, while 
$R^{\a}_{~\b\g\d}$,
$R_{\a\b}$ and $R=R_{\a\b}g^{\a\b}$ are the Riemann tensor, 
the Ricci tensor and the scalar curvature, 
respectively\footnote{Our
conventions are as follows :  
the metric has signature $(-,+,\ldots,+)$,
the Riemann tensor is defined by 
$R^{\a}_{~\b\g\d}=\pa_{\g}\G^{\a}_{~\b\d}+\G^{\a}_{~\g\l}\G^{\l}_{~\b\d}
-(\g\leftrightarrow\d)$ and the Ricci tensor
is given by $R_{\m\n}=R^{\a}_{~\m\a\n}$.}.
This action leads to fourth order differential equations and involves no 
dimensional coupling constant. It possesses, together with its 
supersymmetric extensions, remarkable properties (see \cite{FT}
for a review of conformal gravity and conformal supergravity).

In the linearized limit, (\ref{actionconf}) reduces to
\be
S_0[h_{\m\n}]=\frac{1}{2}\int d^4x {\cal W}_{\a\b\g\d} {\cal W}^{\a\b\g\d},
\label{freeactionconf}
\ee
for $g_{\m\n}=\h_{\m\n}+ \alpha
h_{\m\n}$, $h_{\m\n}$ and $\alpha$ being dimensionless.
Here, $ {\cal W}^{\a}_{~\b\g\d}$ is the linearized Weyl tensor
constructed out of the $h_{\m \n}$ according to formula
(\ref{linearizedWeylbis}) below.

The free action (\ref{freeactionconf}) is invariant under both linearized 
diffeomorphisms and the linearized version of the Weyl rescalings,
\be
\d_{\h,\f}h_{\m\n}=\pa_{\m}\h_{\n}+\pa_{\n}\h_{\m}+2\f \h_{\m\n}.
\label{transfolin}
\ee

This paper investigates the extent to which the complete Weyl action 
(\ref{actionconf}) follows from the free action (\ref{freeactionconf}) and the 
requirement of consistent self-interactions. We show that (\ref{actionconf})
is in fact the only way -under precise assumptions which will be spelled out 
in detail below- to make the fields $h_{\m\n}$ self-interact given the starting
point  (\ref{freeactionconf}). In the same manner, the gauge transformations
\be
\frac{1}{\alpha} \d_{\h,\f}g_{\m\n}=\h_{\m;\n}+\h_{\n;\m}+2\f g_{\m\n}
\label{transfononlin}
\ee
constitute the only possibility to deform the abelian gauge symmetry 
(\ref{transfolin}) (in a way compatible with the existence of a variational
principle that reduces to (\ref{freeactionconf}) in the free limit).

The conditions under which these conditions hold are
\begin{itemize}
 \item
   the deformation is smooth in a formal deformation parameter $\alpha$
   and reduces, in the free limit where $\alpha=0$, to the original theory;
 \item
   the deformation preserves Lorentz invariance;
 \item
   the deformed action involves at most four derivatives of $h_{\m\n}$.
\end{itemize}
This latter condition is equivalent to the requirement that the coupling 
constants be dimensionless or of positive mass dimension. Note that since
 the field $h_{\m\n}$ is dimensionless, the space of dimension-4 polynomials
in $h_{\m\n}$ and its derivatives is infinite-dimensional.

As announced,
our approach is based on the BRST-antifield formalism 
\cite{ZinnJust,dWvH,BV,M1990,GPS1995,BOOK}
where consistent
deformations of the action appear as deformations of the solution of the
master equation that preserve the master equation 
\cite{Barn-Henn,Henneaux:1997bm,Stasheff:1997fe}. 
In that view, the first-order consistent deformations define cohomological
class of the BRST differential at ghost number zero. 
We shall follow closely 
the procedure of \cite{BDGH}, where BRST techniques
were used in the context of the linearized Einstein theory.
A crucial role is played in the calculation by the so-called invariant 
characteristic cohomology, which, just as in the Yang-Mills case 
\cite{Barnich:1995db,Barnich:1995mt,review} or 
in the Pauli-Fierz theory \cite{BDGH}, 
controls the antifield-dependence of the BRST 
cocycles. 
We refer to \cite{Barn-Henn,Henneaux:1997bm,Stasheff:1997fe} 
for background material on the BRST approach to the problem of
consistent deformations.                
The application of BRST techniques to the conformal gravity
context should also prove useful in the investigation of the
cohomological questions raised in \cite{S-T}.

It should be stressed that the requirement of
(background) Lorentz invariance is the only invariance requirement imposed
on the interactions.  There is no a priori condition of diffeomorphism
invariance or Weyl invariance.  These automatically follow from the 
consistency conditions.  No condition on the polynomial degree of the
first order vertices or of the gauge transformation is imposed either.
Finally, it is not required that the deformed gauge symmetries should
close off-shell.  This also follows automatically from the consistency
requirement.  In fact, the structure of the deformed gauge algebra is
severely constrained even if one does not impose conditions on
the number of derivatives, see section \ref{A3} below.
Off-shell closure is automatic ({\it to
first-order in the deformation parameter}) no matter how many
derivatives one allows in the interactions. 
Furthermore, there are only three possible deformations
of the gauge algebra, two of them involving more derivatives than
(\ref{transfononlin}) and being excluded if one imposes the third condition
above. 

\subsection{Interactions for a collection of Weyl gravitons}
Another remarkable feature of Einstein theory, besides the uniqueness
of the graviton vertex, is that it involves a single
type of gravitons (in contrast to Yang-Mills theory, which involves
a collection of 
spin-1 carriers of the YM interactions). As discussed recently
in \cite{BDGH}, this is not an accident.
Namely, the only consistent deformations of the free action
\be
S_0^{PF} [h_{\m\n}^{a}] = \sum_{a=1}^N S_0^{PF} [h_{\m\n}^{a}]
\label{sumPaFi}
\ee
for a collection of massless spin-2 fields, involving no more derivatives than
the free action, cannot introduce true cross-couplings between the different
types of gravitons: by a change of basis in the internal space
of the gravitons, one can always remove any such cross-interactions. 

One may wonder whether a similar property holds in the
conformal case.  We have thus also
investigated the deformations
of the free action
\be
S_0 [h_{\m\n}^a] = \frac{1}{2}\sum_{a=1}^N 
\int d^4x {\cal W}^a_{\a\b\g\d} {\cal W}^{a \a\b\g\d}
\label{many}
\ee
describing a collection of free Weyl gravitons.
We have found that there is again no unremovable cross-interactions 
that can be introduced among the Weyl gravitons.  A key
role is played in the derivation of this result by the
requirement that the action should reduce to (\ref{many})
in the free limit.  If, instead of (\ref{many}), one allows
a free action in which the positive definite metric
$\delta_{ab}$ in the internal space of the Weyl gravitons
is replaced by a metric
$k_{ab}$ with indefinite signature, then cross-interactions
become possible.  We give an explicit example in the paper.
In the case of ordinary gravitons, it is meaningful to
request that all the gravitons should come with the same
sign in the action (\ref{sumPaFi}) since otherwise, the 
energy would be unbounded from below.  However, in the
conformal case, the energy is in any case unbounded
already for a single Weyl graviton
so the legitimacy of
this requirement does not appear to be as clear
as in the standard case.

\subsection{Outline of paper}

The paper is organized as follows. In the next section, we set up our 
notations and formulate precisely the cohomological question. We then compute 
the cohomology of the differential $\g$ related to the gauge transformations
(\ref{transfolin}) (section 3). 
In section 4 we apply standard cohomological results to
compute various cohomologies in the particular case of linearized Weyl
gravity.
Section 5 is the core of out paper. We calculate the invariant characteristic 
cohomology $H^{inv}(\d \vert d)$ in form degree  $n$ and and antifield number
$\geq$ 2. Our computation is quite general in that no restriction on the
dimensionality of the cycles is imposed at this stage. The calculation of
$H^{inv}(\d \vert d)$ 
follows the general pattern developed previously for the 
Yang-Mills field \cite{Barnich:1995db,Barnich:1995mt}
or the massless spin-2 field \cite{BDGH}, but present additional 
novel features related to the algebraic nature of the Weyl symmetry
(no derivative of gauge parameters). 
The relevance of $H^{inv}(\d \vert d)$ 
for the problem of consistent interactions appears clearly in section 6, where
we derive all the consistent deformations fulfilling the above conditions
of Lorentz invariance and dimensionality. We show that there is in fact a 
unique deformation, given by the ${\co} (\a)$-term in the action 
(\ref{actionconf}). Uniqueness to all orders is then easily established.
The analysis is carried out up to this stage with a single
symmetric field $h_{\m\n}$.  In section 7, we discuss deformations
for the action (\ref{many}) involving many symmetric fields
$h_{\m\n}^a$ and show that cross-interactions are impossible 
(in four dimensions) if one insists that the free action be
indeed (\ref{many}), but that cross-interactions can be introduced
if one allows a free limit in which some of the Weyl gravitons
are described by an action with the opposite sign.
Section 7 is devoted to the conclusions and a brief discussion
of the extension of our results to other spacetime dimensions $\geq 4$
as well as $n=3$.
The case $n=2$ has been studied in \cite{BVPT1,BVPT2}.

%%%%%%%%%%%%%%%%%%%%%%%%%%%%%%%%%%%%%%%%%%%%%%%%%%%%%
\section{BRST-symmetry and conventions}
%%%%%%%%%%%%%%%%%%%%%%%%%%%%%%%%%%%%%%%%%%%%%%%%%%%%%
\setcounter{equation}{0}  
\setcounter{theorem}{0}  
\setcounter{lemma}{0} 
\subsection{Differentials $\delta$, $\gamma$ and $s$}
We start the calculation in all spacetime dimensions $\geq 3$.
The restriction $n=4$ will be imposed only at the very end.
Similarly, we do not impose Lorentz invariance
or any condition on the derivative order.  These will not be needed
before section 6.
By following the general prescriptions of the antifield formalism
\cite{BV,BOOK}, one finds that the spectrum of fields, ghosts and
antifields, with their respective gradings, is given by
\bqn
\begin{array}{c|c|c|c}
Z & \mathit{puregh}(Z) & \mathit{antigh}(Z) & \mathit{gh}(Z)
\\
\hline\rule{0em}{3ex}
h_{\m\n} & 0 & 0 & 0
\\
\rule{0em}{3ex}
C_{\m} & 1 & 0 & 1
\\
\rule{0em}{3ex}
\x & 1 & 0 & 1
\\
\rule{0em}{3ex}
h^{*\m\n} & 0 & 1 & -1
\\
\rule{0em}{3ex}
C^{*\mu} & 0 & 2 & -2
\\
\rule{0em}{3ex}
\x^* & 0 & 2 & -2
\end{array}
\eqn                  
where $C_{\m}$ are the ghosts for the diffeomorphisms and $\x$
the ghost corresponding to the Weyl transformation. The variables
$h^{*\m\n}$, $C^{*\mu}$ and $\x^*$ are the antifields.
The antighost number is also called the antifield number and we will use 
both terminologies here.
The BRST-differential for the linear theory (\ref{freeactionconf}),
(\ref{transfolin}) is given
by
\be
s=\d+\g
\label{s}
\ee
where $\d$ is the Koszul-Tate differential and $\g$ is the exterior derivative
along the gauge orbits. Explicitly,
\bqn
&&\g h_{\m\n}=2\pa_{(\m}C_{\n)}+2\h_{\m\n}\x; {~~~~~}\g\x=\g C_{\n}=0;
\nonumber \\
&&\g h^{*\m\n}=0=\g C^{*\m}=\g \x^*;
 \label{gamma} \\
&&\d h_{\m\n}=0, ~~~~\d C^{*\m}=-2\pa_{\n}h^{*\m\n};
                          ~~~\d \x^*=2\h_{\m\n}h^{*\m\n};
\nonumber \\
&&\d h^{*\m\n}=\frac{\d {\cal L}_0}{\d h_{\m\n}}.
\label{definitions}
\eqn      
where ${\cal L}_0$ is the free Lagrangian.
In $n$ dimensions, the free Lagrangian contains
$n$ derivatives of $h_{\m \n}$ (more on its structure
in the conclusions).
The BRST-differential $s$ raises the ghost number and $\g$ raises the 
pureghost 
number by one unit, while $\d$ decreases the antighost number by one unit.

Using the  relations (\ref{gamma}), (\ref{definitions})
and the Noether identities for
the Euler-Lagrange derivatives of ${\cal L}_0$
(i.e., $\partial_\m (\frac{\d {\cal L}_0}{\d h_{\m\n}}) = 0$,
$\frac{\d {\cal L}_0}{\d h_{\m\n}} \eta_{\m \n} = 0$), together
with the invariance of these Euler-Lagrange derivatives under (\ref{transfolin}),
it is easy to check that 
\be
\d^2=0,~~~\g^2=0, ~~~\d \g+ \g \d =0.
\ee
Hence,
\be
s^2 = 0.
\ee

\subsection{Consistent deformations and cohomology}
The solution $\stackrel{(0)}{W}$ of the master equation for the free
theory 
\be
(\stackrel{(0)}{W},\stackrel{(0)}{W})=0,
\ee 
is
\be
\stackrel{(0)}{W} = S_0+2\int d^n x h^{*\m\n}(\pa_{\m}C_{\n}+\h_{\m\n}\x). 
\ee
The BRST differential $s$ is related to $\stackrel{(0)}{W}$ through
\be
sA \equiv (\stackrel{(0)}{W},A).
\ee
As it is well known, the solution of the master equation
captures all the information about the gauge invariant action,
the gauge symmetries and their algebra. The BRST approach
to the problem of consistent interactions consists in deforming
$\stackrel{(0)}{W}$,
\be
\stackrel{(0)}{W} \rightarrow W = \stackrel{(0)}{W} + \alpha \stackrel{(1)}{W}
+ {\cal O}(\alpha^2)
\ee
in such a way that the master equation is fulfilled to each order
in $\alpha$ \cite{Barn-Henn,Henneaux:1997bm}
\be
(W,W) = 0.
\ee
This guarantees gauge invariance by construction
(this is what we mean by ``consistency" in this paper: the deformed
action is ``consistent" if and only
if it is invariant under a set of
gauge transformations which reduce to (\ref{transfononlin}) in the free limit).

One then proceeds order by order in $\alpha$.  The first-order deformations
$\stackrel{(1)}{W}$ must fulfill
\be
(\stackrel{(0)}{W}, \stackrel{(1)}{W}) \equiv  s \stackrel{(1)}{W} = 0.
\ee
Trivial solutions of this BRST-cocycle condition,
of the form $sK$ for some $K$, correspond to trivial deformations
that can be undone by a change of variables.  Thus, the non-trivial first-order
deformations are described by the cohomology group $H^0(s,{\cal F})$
of the BRST differential in the space ${\cal F}$ of local 
functionals, or, what is the same, the group $H^{0,n}(s\vert d)$
of the BRST differential acting in the space of local
$n$-forms ($0$ = ghost number, $n$ = form-degree).  The local
$n$-forms are by definition given by
\be
\omega = f([h_{\m\n}],[C_\m],[\xi],[h^{*\m\n}],[C^{*\m}],[\xi^{*}])
\; \; dx^0 \wedge \cdots \wedge dx^{n-1}
\ee
where $f$ is a function of $h_{\m\n}$, $C_\m$, $\xi$, $h^{*\m\n}$,
$C^{*\m}$, $\xi^{*}$ and their derivatives up to some finite (but unspecified)
order (``local function").  This is what is meant by the notation
$f([h_{\m\n}],[C_\m],[\xi],[h^{*\m\n}],[C^{*\m}],[\xi^{*}])$.
Thus, a ``local function'' $f$ of $\phi$, $f=f([\phi])$, is a function of
$\phi$ and a finite number of its derivatives.  In fact,
we shall assume that $f$ is polynomial in all the variables except, possibly,
$h_{\m\n}$.

The knowledge of  $H^0(s\vert d)$ requires the computation of the following
cohomological groups : $H(\g)$, $H(\g \vert d)$, $H(\d)$,  $H(\d \vert d)$
and  $H^{inv}(\d \vert d)$ (see \cite{Barnich:1995db,Barnich:1995mt}). 
Our first task is therefore to compute these groups.

%%%%%%%%%%%%%%%%%%%%%%%%%%%%
\section{Cohomology of $\g$}  
%%%%%%%%%%%%%%%%%%%%%%%%%%%%
\setcounter{equation}{0}  
\setcounter{theorem}{0}  
\setcounter{lemma}{0} 
In this section we calculate explicitly 
\be
H(\g) \equiv \frac{Ker (\g)}{Im (\g)}.
\ee
For that purpose, it is convenient to
split $\g$ as the sum of the operator $\g_{0}$
associated with linearized diffeomorphisms plus
the operator $\g_{1}$ associated with
linearized Weyl transformations :
\be
\g=\g_{0}+\g_{1}.
\ee      
The grading associated to this splitting is the number of
ghosts $\xi$ and their derivatives ($\g_{1}$ increases this
number by one unit, while $\g_{0}$ does not affect it).

\subsection{Linearized conformal invariants}
First, we recall a few known facts.  Let $f([h_{\mu \nu}])$ be
a local function of $h_{\mu \nu}$.
This function is invariant under linearized
diffeomorphisms if and only if it involves only the
linearized Riemann tensor ${\cal R}_{\a \b \m \n}$ and its derivatives,
\be
\g_{0} f = 0 \Leftrightarrow f = f([{\cal R}_{\a \b \m \n}])
\ee
with\footnote{The components of the linearized Riemann tensor 
and their derivatives are not
independent because of the linearized Bianchi identities but this does
not affect the argument - it simply implies that there are many ways to
write the same function.  One could choose a set of independent
components and work exclusively with this set, but this will not
be necessary for our purposes.}
\be
{\cal R}_{\a \b \m \n} = -\frac{1}{2}(\pa_{\a\m}h_{\b\n}+\pa_{\b\n}h_{\a\m}
-\pa_{\a\n}h_{\b\m}-\pa_{\b\m}h_{\a\n}).
\ee
To discuss Weyl invariance, one
introduces the linearized Weyl tensor ${\cal W}_{\a \b \m \n}$,  given
by 
\be
{\cw}_{\a \b \g \d} = {\cal{R}}_{\a \b \g \d}
-\frac{2}{n-2}(\h_{\a[\g}{\cal{R}}_{\d]\b}-\h_{\b[\g}{\cal{R}}_{\d]\a})
+\frac{2}{(n-2)(n-1)}\cal{R} \h_{\a[\g}\h_{\d]\b},
\label{linearizedWeylbis}
\ee
where 
\be
{\cal{R}}_{\m\n}=\frac{1}{2}\h^{\a\l}
(h_{\l\n,\m\a}-h_{\m\n,\l\a}-h_{\l\a,\m\n}+h_{\m\a,\l\n}),
\ee
and
\be
{\cal{R}}=\h^{\a\l}\h^{\m\n}(h_{\m\a,\l\n}-h_{\m\n,\a\l}).
\ee
The symmetries of the Weyl tensor are
${\cw}_{\a\b\g\d}=-{\cw}_{\b\a\g\d}=-{\cw}_{\a\b\d\g}={\cw}_{\g\d\a\b}$
as well as the cyclic identity ${\cw}_{\a[\b\g\d]}=0$.
All the traces of ${\cw}_{\a\b\g\d}$ vanish.
One also introduces the linearization ${\cal K}_{\a \b}$ of
the tensor $K_{\a \b}$ defined in the introduction,
\be
\ck_{\m\n}=\frac{1}{n-2}\Big[{\cal{R}}_{\m\n}-\frac{1}{2(n-1)}
\h_{\m\n}{\cal{R}}\Big].
\ee
One has
\be 
{\cw}_{\a \b \g \d} = {\cal{R}}_{\a \b \g \d}  - 2 
(\eta_{\a [\g}{\cal K}_{\d]\b} - \eta_{\b [\g} {\cal K}_{\d]\a}).
\label{decompRiem}
\ee
The linearized Weyl tensor ${\cal W}_{\a \b \m \n}$ is annihilated
by $\g_{1}$, 
\be
\g_{1} {\cal W}_{\a \b \m \n} = 0,
\ee
while one has
\be
\g_{1} {\cal K}_{\a \b} = - \partial^2_{\a \b} \xi.
\label{g1K}
\ee
The (linearized) Cotton tensor ${\cal C}_{\a \b \m}$ is obtained by taking the
antisymmetrized derivatives of ${\cal K}_{\a \b}$,
\be
{\cal C}_{\a \b \m} = 2\pa_{[\m}\cal K_{\b]\a}
\ee 
and is clearly Weyl-invariant
\be
\g_{1} {\cal C}_{\a \b \m} = 0.
\ee
The linearized Bianchi identities imply
\be
\partial_\a {\cal W}^\a_{\; \; \b \m \n} =
(3-n) {\cal C}_{\b \m \n}.
\ee
In $n \geq 4$ spacetime dimensions, the Cotton tensor
is therefore not independent from the derivatives of the Weyl tensor.  Any
local function of the Riemann tensor can be expressed as a local
function of the Weyl tensor and the symmetrized derivatives of
${\cal K}_{\a \b}$,
\be
f([{\cal R}_{\a \b \m \n}]) \sim f([{\cal W}_{\a \b \m \n}], {\cal K}_{\a \b},
\partial_{(\m} {\cal K}_{\a \b)}, 
\partial^2_{(\m \n}{\cal K}_{\a \b)}, \cdots). 
\ee
Because of (\ref{g1K}), 
it is invariant under (linearized) Weyl transformations if and only if
it depends only on the Weyl tensor and its derivatives,
\be
\g_{1} f([{\cal R}_{\a \b \\m \n}]) = 0 \Leftrightarrow f = 
f([{\cal W}_{\a \b \m \n}])\; \; \; (n \geq 4).
\ee

In three spacetime dimensions, the Weyl tensor identically vanishes.
Any local function of the Riemann tensor can be expressed as a local
function of the tensor ${\cal K}_{\a \b}$, or, what is the same, as a local function
of the Cotton tensor and the symmetrized derivatives of
${\cal K}_{\a \b}$,  
\be
f([{\cal R}_{\a \b \m \n}]) \sim f([{\cal C}_{\a \b \m}], {\cal K}_{\a \b},
\partial_{(\m} {\cal K}_{\a \b)}, \partial^2_{(\m \n}{\cal K}_{\a \b)}, \cdots).
\ee
It is invariant under (linearized) Weyl transformations if and only if
it depends only on the Cotton tensor and its derivatives,
\be
\g_{1} f([{\cal R}_{\a \b \m \n}]) = 0 \Leftrightarrow f = f([{\cal C}_{\a \b \m}])
\; \; \; (n = 3).
\ee 
Note that the Cotton tensor is equivalent to its ``dual''
defined by
$\cc_{\a\b}=\frac{1}{2}\h_{\a\r}\varepsilon^{\r\l\m}\cc_{\b\m\l}$
(in 3 dimensions). This 
invariant tensor 
$\cc_{\a\b}=\pa^{\m}[(-)\varepsilon_{\m\a\n}\ck^{\n}_{~\b}]$
is symmetric by virtue of the Bianchi identity 
and clearly fulfills $\pa_{\a}\cc^{\a\b}=0$ as well as $\cc^{\a\b}\h_{\a\b}=0$.

\subsection{The Bach tensor}

Of particular importance in $4$ spacetime dimensions is the (linearized)
Bach tensor \cite{bach},
\be
{\cal B}^{\a \b} = (-2)\pa_{\r}\cal C^{\a\b\r} 
\label{linbach}
\ee
which is such that
\be
\frac{\delta {\cal L}_0}{\delta h_{\a \b}} = {\cal B}^{\a \b}
\; \; \; (n=4).
\ee
The linearized Bach tensor is symmetric (by virtue of the Bianchi identities),
 gauge-invariant, Lorentz-covariant, traceless and divergencefree,
\be
{\cal B}^{\a \b} = {\cal B}^{\b \a}, \; \;
\g {\cal B}^{\a \b} = 0, \; \; \partial_\a {\cal B}^{\a \b} = 0, \; \;
{\cal B}^{\a}_{\;\; \a} = 0.
\ee
These last two properties are direct consequence of the definition 
(\ref{linbach}) where we stress that $\cc^{\a\b\r}=-\cc^{\a\r\b}$, 
$\cc^{\a\b\r}\h_{\a\b}=0$.
Actually, the Bach tensor
is the only tensor containing four derivatives (or less)
of $h_{\a\b}$ with these properties.
Indeed, such a tensor must be obtained by contracting indices
in $\pa_\rho \pa_\sigma {\cal W}_{\a \b \g \d}$,
${\cal W}_{\a \b \g \d} {\cal W}_{\l \m \n \rho}$ (four
derivatives), $\pa_\rho {\cal W}_{\a \b \g \d}$ (three derivatives)
or ${\cal W}_{\a \b \g \d}$ (two derivatives).  The last two
possibilities are clearly excluded.  There is only one independent
way to contract indices in $\pa_\rho \pa_\sigma
{\cal W}_{\a \b \g \d}$ and this 
produces a tensor proportional to ${\cal B}_{\a  \g}$.  Finally, the
only possible contraction ${\cal W}^{\a}_{\; \; \b \g \d}
{\cal W}^{\l\b \g \d}$ is ruled out because it fails to be traceless.

\subsection{Computation of $H(\g)$}

We want to find the most general solution of the cocycle condition
\be
\g a=0 
\ee
where $a$ has a definite pure ghost number, say $k$.
Let us expand $a$ with respect to the powers of $\x$ and
its derivatives,
\be
 a=a_0+a_1+a_2+\ldots
\ee
where $a_0$ contains neither $\xi$ nor its derivatives, $a_1$
is linear in $\xi$ or one of its derivatives, etc.
The expansion stops at order $k$ equal to the pure ghost number
of $a$ (or earlier).  If one plugs this expansion into $\g a = 0$ with $\g
= \g_0 + \g_1$, one gets
at zeroth order
$\g_0 a_0=0$.  This implies \cite{BDGH}\footnote{The differential $\g_0$ is
just the differential $\g$ of \cite{BDGH}, with the addition of the
``$\g_0$-singlet" variables $\x$, $\x^*$ and their derivatives, which
are annihilated by $\g_0$ and define new independent generators in the
$\g_0$-cohomology. As shown in \cite{BDGH}, the $\g_0$-cocycle condition forces
the cocycles to depend on $h_{\m \n}$ only through
the linearized Riemann tensor and its derivatives, and eliminates
all second and higher derivatives of $C_\m$ as well as the symmetrized
derivatives $\pa_\m C_\n + \pa_\n C_\m$ (up to trivial terms).}
$a_0=P_\Delta([{\cal R}_{\m\n\a \b}], [\Phi^*]) \, 
w^\Delta (C_{\m}, \pa_{[\m}C_{\n]})+\g_0 b_0$,
where 
$\Phi^*$ denotes collectively all the antifields and where
the $\{w^\Delta\}$ form a basis of polynomials in $C_{\m}$,
$\pa_{[\m}C_{\n]}$.
By the trivial redefinition $a$ $\rightarrow$ $a-\g b_0$ we can set
$a_0=P_\Delta([{\cal R}_{\m\n\a \b}], [\Phi^*]) \, 
w^\Delta(C_{\m}, \pa_{[\m}C_{\n]})$.

At first order, the cocycle condition
reads $\g_1 a_0+\g_0 a_1=0$, that is, 
$\g_1 P_\Delta([{\cal R}_{\m\n\a \b}], [\Phi^*])$
$w^\Delta(C_{\m}, \pa_{[\m}C_{\n]}) +\g_0 a_1=0$.
But  $\g_1 P_\Delta$ is a polynomial in ${\cal R}_{\m\n\a \b}$,
$\Phi^*$, $\xi$ and their derivatives, which  cannot be $\g_0$-exact unless
it vanishes \cite{BDGH} ($\xi$ and its derivatives are $\g_0$-closed but
not $\g_0$-exact).
Thus, one gets
$\g_1 P_\Delta=0$, which implies 
$P_\Delta=P_\Delta([{\cal W}_{\m\n\a \b}], [\Phi^*])$ 
($n \geq 4$) or $P_\Delta=P_\Delta([{\cal C}_{\m\n\a }], [\Phi^*])$
($n=3$).  In the sequel, we shall assume for definiteness that
$n \geq 4$. 
Thus $a_0= P_\Delta([{\cal W}_{\m\n\a \b}], [\Phi^*]) \, 
w^\Delta(C_{\m}, \pa_{[\m}C_{\n]})$.
There remains $\g_0 a_1=0$, which implies 
$a_1=Q_\Delta([{\cal R}_{\m\n\a \b}],[\x], [\Phi^*]) \, 
w^\Delta(C,\pa_{[\m} C_{\n]})+\g_0 b_1$.
We redefine $a~\rightarrow a-\g b_1$ to eliminate the $\g_0 b_1$ term
from $a_1$.

At second order $\g_1 a_1+\g_0 a_2=0$ reads 
$\g_1 Q_\Delta([{\cal R}_{\m\n\a \b}],[\x], [\Phi^*]) \, 
w^\Delta(C,\pa_{[\m} C_{\n]})+\g_0 a_2 = 0$.
As before, $\g_1Q_\Delta$ cannot be $\g_0$-exact unless it vanishes, 
thus $\g_1 Q_\Delta=0$.  
The solve this, we note that the cohomology of $\g_1$ in the
space of functions $f([{\cal R}_{\m\n\a \b}],[\x], [\Phi^*])$ is
easily evaluated since the tensor ${\cal K}_{\a \b}$ and its
successive symmetrized derivatives form trivial pairs with the
derivatives $\pa_{\mu_1 \cdots \mu_\ell} \xi$,  $\ell \geq 2$.  Thus,
only ${\cal W}_{\m\n\a \b}$ and its derivatives, $\xi$ and $\pa_\m \xi$ 
remain in cohomology.
This implies 
that $Q_\Delta=Q_\Delta([{\cal W}_{\m\n\a \b}],[\Phi^*], \x, \pa_{\m}\x)$
$+\g_1K$ with some $K$ that involves the $h_{\m \n}$
only through the curvatures, {\it{i.e.}} is such that $\g_0 K = 0$.
Therefore, $a_1$ reads
\\ 
$a_1=Q_I([{\cal W}_{\m\n\a \b}],[\Phi^*]) \, 
\o^I (C_\m ,\pa_{[\m}C_{\n]}, \x,\pa_\m \x) + \g_1 K$ and one
can remove $\g_1 K$ by a trivial redefinition.
Here, the $\{\o^I (C_\m ,\pa_{[\m}C_{\n]}, \x,\pa_\m \x) \}$
form a basis of polynomials in $C_\m$, $\pa_{[\m}C_{\n]}$,
$\x$ and $\pa_\m \x$.
The condition for $a_2$ becomes then $\g_0 a_2=0$ 
and the procedure continues to all orders in powers of
the Weyl ghost and its derivatives.
\\
{\bf{Conclusion}} : the cohomology of $\g$ is
isomorphic to the space of functions of ${\cal W}_{\m\n\a \b}$,
$\Phi^*$, their derivatives, $C_{\m}$, $\pa_{[\m}C_{\n]}$,
$\x$ and $\pa_\m \x$,
\be
H(\g) \simeq \left\{ f([{\cal W}_{\a\b\g\d}], [\Phi^*], \x, \pa_{\m}\x, C_{\m},
   \pa_{[\m}C_{\n]}) \right\} \; \; \; (n \geq 4).
\ee
In three dimensions, 
\be
H(\g) \simeq \left\{ f([{\cal C}_{\a\b\g}], [\Phi^*], \x, \pa_{\m}\x, C_{\m},
   \pa_{[\m}C_{\n]}) \right\} \; \; \; (n = 3).
\ee  
Note that the ghost derivatives  $\x$, $\pa_{\m}\x$, $C_{\m}$
and $\pa_{[\m}C_{\n]}$ that survive in cohomology are in number equal to the
number of infinitesimal transformations of the conformal group.
In pure ghost number zero,
the polynomials $\a([{\cal W}_{\m\n\a \b}],[\Phi^*])$ 
(or $\a([{\cal C}_{\m\n\a }],[\Phi^*])$) 
are called "invariant polynomials".

%%%%%%%%%%%%%%%%%%%%%%%%%%%%%%%%%%%%%%%%%%%
\section{Standard material : $H(\g \vert d)$, $H(\d)$, $H(\d \vert d)$}
%%%%%%%%%%%%%%%%%%%%%%%%%%%%%%%%%%%%%%%%%%%
%
\setcounter{equation}{0}  
\setcounter{theorem}{0}  
\setcounter{lemma}{0} 
%~~~~~~~~~~~~~~~~~~~~~~~~~~~~~~~~~~~~~~~~
\subsection{General properties of $H(\g \vert d)$}
%~~~~~~~~~~~~~~~~~~~~~~~~~~~~~~~~~~~~~~~~
\label{A1}

Now that we know $H(\g)$, we can consider $H(\g \vert d)$, the space of 
equivalence classes of forms
$a$ such that $\g a+db=0$, identified by the relation $a\sim a'$ 
$\Leftrightarrow$ $a'=a+\g c+df$.
We shall need properties of $H(\g \vert d)$ in strictly
positive antighost (= antifield) number.  To
that end, we first recall the following theorem on invariant polynomials
(pure ghost number $=0$) :
\begin{theorem}\label{2.2}
In form degree less than n and in antifield number strictly greater than
$0$, the cohomology of $d$ is trivial in the space of invariant polynomials.
\end{theorem}
{\bf{Proof}}: This  just follows from the standard algebraic
 Poincar\'e lemma in the sector of the antifields \cite{Duboisetal}.

Theorem \ref{2.2}, which deals with $d$-closed
invariant polynomials that involve no
ghosts, has the following useful consequence on 
general mod-$d$ $\g$-cocycles with $antigh >0$. 
\\
{\bf{Consequence of Theorem \ref{2.2}}}

{\it{If $a$ has strictly positive antifield number
(and involves possibly the ghosts), the
equation
\begin{equation}  
\gamma a + d b = 0  
\end{equation}  
is equivalent, up to trivial redefinitions, to  
\begin{equation}  
\gamma a = 0 . 
\end{equation}  
That is, one can add $d$-exact terms to $a$,  
$a \rightarrow a' = a  + d v$ so that $\g a'= 0$.  
Thus, in antighost number $>0$, one can always
choose representatives of $H(\g \vert d)$ that are
strictly annihilated by $\g$}}.

This does not imply
that $H(\g \vert d)$ and $H(\g)$ are isomorphic:
although the cocycle conditions are equivalent, the coboundary
conditions are different.  However, this is all that we
shall need about the equation $\gamma a + d b = 0$.
  
In order to prove that $\gamma a + d b = 0$ can be
replaced by $\gamma a = 0$ in strictly positive antifield
number, we consider the  
descent associated with $\gamma a + d b = 0$:  
by using the properties $\gamma^2 = 0$,  
$\gamma d + d \gamma = 0$ and the triviality of the cohomology of $d$
(algebraic Poncar\'e lemma), 
one infers that  
$\gamma b + dc = 0$ for some $c$.  Going on in the same way, we introduce  
the chain of equations $\gamma c + de = 0$, $\gamma e + d f = 0$, etc,  
in which each  
successive equation has one less unit in form-degree.  
The descent ends with the last two equations $\gamma m + dn = 0$,  
$\gamma n = 0$ (the last equation is $\gamma n = 0$ either because $n$  
is a zero-form, or because one stops earlier with a $\gamma$-closed  
term).  
  
Now, because $n$ is $\gamma$-closed, one has, up  
to trivial, irrelevant terms,  
$n = \a_J \omega^J$ where (i) the $\a_J = a_J([{\cal W}_{\a \b \g \d},
[\Phi^*])$ are invariant polynomials;
and (ii) the $\omega^J = \omega^J (\x, \pa_\m \x, C_\m, 
\pa_{[\m} C_{\n ]})$ form, as before,
a basis of the space of polynomials
in $\x$, $\pa_\m \x$, $C_\m$, $\pa_{[\m} C_{\n ]}$.  
The $dx^\m$'s are included in the $\a_J$'s, i.e., the $\omega^J$'s
are $0$-forms.
Inserting this expression into the previous  
equation in the descent yields  
\be  
d (\a_J) \omega^J \pm \a_J d \omega^J + \gamma m = 0 . 
\label{keya3}  
\ee  
In order to analyse (\ref{keya3}), we introduce a new differential $D$,  
whose action on $h_{\mu \nu}$,  
$h^{\star}_{\mu \nu}$, $C^{\star}_{\alpha}$ and all their derivatives is the  
same as the action of $d$, but whose action on the ghosts 
occurring in $\omega^J$ is given by :  
\begin{eqnarray}  
D C_{\mu} &=&  d x^{\nu} C_{[\mu , \nu ]}
-d x^{\nu}\h_{\m\n}\x  
\nonumber \\  
D(\pa_{[\r}C_{\m]}) &=& 2 d x^{\l}\h_{\l[\r}\pa_{\m]}\x 
\nonumber \\
D\x&=&d x^{\nu}\pa_{\n}\x
\nonumber \\
D(\pa_{\m}\x)&=&0.
\end{eqnarray}  
The operator $D$ coincides with $d$ up to $\gamma$-exact terms.   
It follows from the definitions that $D\omega^J = A^J_I \omega^I$  
for some constant matrix $A^J_I$ linear in $dx^\m$.  
  
One can rewrite (\ref{keya3}) as  
\be  
d (\a_J) \omega^J \pm \a_J D \omega^J + \gamma m' = 0  
\ee  
which implies,  
\be  
d (\a_J) \omega^J \pm \a_J D \omega^J = 0  
\label{keya4}  
\ee  
since a term of the form $\b _J \omega^J $ (with $\b _J$  
invariant) is $\g$-exact if and only if  
it is zero.  
It is convenient to further split $D$ as the sum of an operator $D_0$ and an   
operator $D_1$, by assigning a $D$-degree to the variables 
occurring in the $\g$-cocycles $\a_I\o^I$ as follows. 
Everything has $D$-degree zero, except the ghosts and their
derivatives to which we assign :
\bqn
C_{\m} &\rightarrow& D{\rm{-degree}}~ 0,
\nonumber \\
\pa_{[\r}C_{\m]}&\rightarrow& D{\rm{-degree}}~ 1,
\nonumber \\
\x &\rightarrow& D{\rm{-degree}}~ 1,
\nonumber \\
\pa_{\m}\x &\rightarrow& D{\rm{-degree}}~ 2.
\eqn
The $D$-degree is bounded
because there is a finite number of $C_{[\mu , \nu]}$,
$\x$, $\pa_{\m}\x$, which are
anticommuting.    One has
$D= D_0 + D_1$, where
$D_0$ has the same action as $D$ on  
$h_{\mu \nu}$, $h^{\star}_{\mu \nu}$, $C^{\star}_{\alpha}$ and all their  
derivatives, and gives $0$ when acting on the ghosts. $D_1$ gives $0$ when   
acting  on all the variables but the  
ghosts, on which it reproduces the action of   
$D$.  
$D_1$ raises the $D$-degree by one unit, while  
$D_0$ leaves it unchanged. 
  
Let us expand (\ref{keya4}) according to the $D$-degree.  
At lowest order, we get  
\be  
d \a_{J_0} = 0  
\ee  
where $J_0$ labels the $\omega^J$ that contain zero derivatives of 
$C_{\m}$, and no $\x$ ($D\omega^J = D_1 \omega^J $ contains at 
least one derivative of $C_{\m}$ or one $\x$).  
This equation implies, according to Theorem  \ref{2.2}, that  
$\a_{J_0} = d \b _{J_0} + \hbox{ $\g$-exact terms}$,  where $\b _{J_0}$
 is an invariant polynomial.  
Accordingly,  one can write  
\bqn  
\a_{J_0}\omega^{J_0} &=&(d\b_{J_0})\omega^{J_0} +\hbox{ $\g$-exact terms}
\nonumber \\
&=& d(\b_{J_0}\omega^{J_0})\mp \b_{J_0} d\omega^{J_0} 
+ \hbox{ $\g$-exact terms}
\nonumber \\
&=&d(\b_{J_0}\omega^{J_0})\mp \b_{J_0} D_1\omega^{J_0}
+\hbox{ $\g$-exact terms}
\nonumber \\
&=&d(\b_{J_0}\omega^{J_0})\mp \b_{J_0}A^{J_0}_{J_1}\omega^{J_1}
+ \hbox{ $\g$-exact terms}.
\eqn 
As explicitly written, the term $\b_{J_0}A^{J_0}_{J_1}\omega^{J_1}$
has $D$-degree equal to $1$.  
Thus, by adding trivial terms to the last term $n$ in the descent,  
we can assume that $n$ contains no term of $D$-degree $0$.  
One can then successively removes the terms of $D$-degree $1$,  
$D$-degree $2$, etc, until one gets that the
last term $n$ in the descent vanishes.  One then repeats the  
argument for $m$ and the previous terms in the descent until one 
reaches the conclusion   
$b = 0$, i.e., $\g a = 0$, as announced.

%~~~~~~~~~~~~~~~~~~~~~~~~~~~~~~~~~~~~~~~~~~~~~~~~~~~~~~~~~~~~~~
\subsection{Characteristic cohomology $H(\d \vert d)$}  
%~~~~~~~~~~~~~~~~~~~~~~~~~~~~~~~~~~~~~~~~~~~~~~~~~~~~~~~~~~~~~~
\label{characteristic}    
We now turn to the cohomological groups involving the Koszul-Tate
differential $\d$. 
A crucial aspect of the differential $\delta$ defined through  
$\d h^{*\a\b}=\frac{\d {\cal L}_0}{\d h_{\a\b}}$, $\d \x^*=2\h_{\m\n}h^{*\m\n}$ 
and $\d C^{*\a}=-2\pa_{\b}h^{*\b\a}$   is that it is related  
to the dynamics of the theory.  This is obvious since  
$\delta h^{*\ m \n}_a$ reproduces the Euler-Lagrange derivatives  
of the Lagrangian.  In fact, one has the following important  
(and rather direct)  
results about the cohomology of $\delta$ \cite{FH,BOOK}  
\begin{enumerate}  
\item Any form of zero antifield number  
which is zero on-shell is $\delta$-exact;  
\item $H^p_i (\delta) = 0$ for $i>0$, where $i$ is the antifield  
number, in any form-degree $p$.    
[The antifield number is written as a lower index; the ghost number is not  
written because it is irrelevant here.]  
\end{enumerate}  
  
We now consider $H(\delta \vert d)$ (known as
the ``characteristic
cohomology" because of an isomorphism theorem established in
\cite{Barnich:1995db} and not needed here).
It has been shown in \cite{locality} that $H(\delta \vert d)$
is trivial in the space of forms with positive pure ghost number. Thus
we need only  $H(\delta \vert d)$ in the space
of local forms that do not involve the ghosts, i.e., having $puregh =0$.   
The following vanishing theorem on $H^n_p(\delta \vert d)$ 
can be proven:  
\begin{theorem}  
\label{vanishing}  
The cohomology groups $H^n_p(\delta \vert d)$ vanish  
in antifield number strictly greater than $2$,  
\be  
H^n_p(\delta \vert d) = 0 \, \hbox{ for } p>2.  
\ee  
\end{theorem}  
The proof of this theorem  
is given in \cite{Barnich:1995db} and follows from the fact that  
linearized conformal gravity is a linear, irreducible, gauge theory.  
 
In antifield number two, the cohomology is given by the following theorem :
\begin{theorem}  
\label{conservation2}  
A complete set of representatives of $H^n_2(\d \vert d)$ is given
by the antifields $C^{*\m}$ conjugate to the diffeomorphism ghosts,
{\it{i.e.}}, 
\be
\d a^n_2+d a^{n-1}_{1}=0 \Rightarrow a^n_2=\l_{\m}C^{*\m}
dx^0\wedge dx^1\wedge\ldots\wedge dx^{n-1}+\d b_3^n+d b_2^{n-1}
\ee
where the $\l_{\m}$ are constants.
\end{theorem}  
  
\noindent  
{\bf{Proof of Theorem \ref{conservation2}}} :   
Let $a$ be a solution of the cocycle condition for $H^n_2(\delta |d)$,  
written in dual notations,  
\be  
\delta a + \partial_\m V^\m = 0.  
\label{coca}  
\ee  
Without loss of generality, one can assume that $a$ is linear in  
the undifferentiated antifields, since the derivatives of $C^{* \m}$  
or $\x^*$
can be removed by integrations by parts (which leaves one in the same  
cohomological class of $H^n_2(\delta |d)$).  Thus,  
\be  
a = \l_\m C^{* \m} + \l \x^* + \m  
\label{exprfora}  
\ee  
where $\m$ is quadratic in the antifields $h^{* \m \n}$ and their  
derivatives, and where the $\l_\m$ and $\l$ can be functions of $h_{\m \n}$  
and their derivatives.
Because $\delta \m \approx 0$, the  
equation (\ref{coca}) implies the linearized conformal Killing equations for  
$\l_{\m}$ and $\l$,  
\be  
\partial_\n \l_\m + \partial_\m \l_\n -2\h_{\m\n}\l \approx 0.  
\label{linearKilling}  
\ee 
To solve this equation, we observe that it
implies
\be
\pa_\a \pa_\n \l_\m \approx \eta_{\m \a} \pa_\n \l
+ \eta_{\m \n} \pa_\a \l - \eta_{\a \n} \pa_\m \l,
\ee
and 
\be
\pa_\a \pa_\b \l \approx 0.
\ee
The second of these equations yields $\pa_\b \l \approx b_\b$ for
some constant $b_\b$ (since $\pa_\m f \approx 0$ implies
$f \approx C$, see \cite{Barnich:1995db}), from which one
gets
\be
\pa_\a \pa_\n \l_\m \approx \eta_{\m \a} b_\n 
+ \eta_{\m \n} b_\a - \eta_{\a \n} b_\m.
\ee
Defining 
\be
\l_\m = \l'_\m + b_{\m}x^2
-2x_{\m}b^{\n}x_{\n}\, , \; \; \l = \l' + b_\m x^\m ,
\ee
we get for $\l'_m$ and $\l'$ the same conformal Killing
equation (\ref{linearKilling}) but now $\pa_\a \pa_\n \l'_\m \approx 0$
and $\pa_\a \l' \approx 0$.  This last equation gives $\l' \approx f$
where $f$ is a constant.  Redefining 
\be
\l'_\m =  \l''_\m + f x_{\m} \, , \; \; \l' = \l'' + f,
\ee
yields the weak Killing equation for $\l''_\m$
\be
\partial_\n \l''_\m + \partial_\m \l''_\n  \approx 0.
\label{weakKilling}
\ee                      
together with $\l'' \approx 0$.  The general solution of (\ref{weakKilling})
is $\l''_\m \approx a_{\m}+\o_{\m}^{~\n}x_{\n}$ with $\o_{\m\n}=-\o_{\n\m}$
(see \cite{BDGH}).  Putting everything together, one gets
\be
\l_{\m}(x) \approx a_{\m}+\o_{\m}^{~\n}x_{\n}+f x_{\m}+b_{\m}x^2
-2x_{\m}b^{\n}x_{\n} \, , \; \; \l \approx f + b_\m x^\m,
\label{solutionWK}
\ee
i.e., $\l_{\m}$ is on-shell equal to a conformal Killing vector
(as expected).  The parameters
$\o_{\m\n}$, $a_{\m}$, $f$ and $b_{\m}$ describe respectively
infinitesimal Lorentz transformations, 
translations, dilations and  
so-called ``special conformal transformations".

We are interested in solutions of (\ref{linearKilling})
that do not depend explicitly on $x^\m$ (since we do
not want the Lagrangian to have an explicit $x$-dependence).  This forces
$\o_{\m}^{~\n} = f = b_{\m} = 0$ 
in (\ref{solutionWK}) and gives
\be
\l_{\m} \approx a_{\m} \, \; \; \l \approx 0.
\ee
Substituting this expression into (\ref{exprfora}), and  
noting that the term proportional to the equations of motion can be  
absorbed through a redefinition of $\m$, one gets  
\be  
a = \lambda_\m  C^{* \m} + \m'  
\label{exprforabis}  
\ee  
(up to trivial terms).  
Now, the first term in the right-hand side of (\ref{exprforabis}) is  
a solution of $\delta a + \partial_\m V^\m = 0$ by itself.  This means that  
$\mu'$, which is quadratic in the $h^{* \m \n}_a$ and their derivatives, must  
be also a $\delta$-cocyle modulo $d$.   But it is well known that  
all such cocycles are trivial \cite{Barnich:1995db}.  Thus,   
$a$ is given by   
\be  
a = \lambda_\m C^{* \m} + {\hbox{ trivial terms}}  
\ee  
where $\lambda_\m$ are constants, as we claimed.  This proves the theorem.  
  
\vspace{.2cm}  
\noindent  
{\bf Comments}   
  
(i)The above theorems provide a complete description of  
$H^d_k(\delta |n)$ for $k>1$. These groups are  
zero ($k>2$) or finite-dimensional ($k=2$).  In  
contrast, the group $H^{n}_1 (\delta |d)$, which is related to  
ordinary conserved currents, is infinite-dimensional since the  
theory is free.  To our knowledge, it has not been completely  
computed.  Fortunately, we shall not need it below.  
(ii)In a recent interesting paper, deformations involving all conformal 
Killing vectors have been investigated \cite{Brandt}.

%%%%%%%%%%%%%%%%%%%%%%%%%%%%%%%%%%%%%%%%%%%%%%%%%%%%%%%%%%
\section{Invariant cohomology of $\delta$ modulo $d$.}  
\label{deltamoddinv} 
%%%%%%%%%%%%%%%%%%%%%%%%%%%%%%%%%%%%%%%%%%%%%%%%%%%%%%%%%%
%
\setcounter{equation}{0}  
\setcounter{theorem}{0}  
\setcounter{lemma}{0} 
\subsection{Central theorem} 
We now establish the crucial result that underlies all our
discussion of consistent interactions for linear Weyl gravity.
This result concerns the  invariant cohomology of $\delta$ modulo $d$.
The group $H^{inv}(\delta \vert d)$ is important because it controls
the obstructions to removing the antifields from a $s$-cocycle
modulo $d$, as we shall see below.

Throughout this 
section, there will be no ghost; i.e., the objects that appear 
involve only the fields, the antifields and their derivatives. 
The central result that gives $H^{inv}(\delta \vert d)$ in
antighost number $\geq 2$ is
\begin{theorem}\label{2.6}  
Assume that the invariant polynomial $a_{k}^{p}$  
($p =$ form-degree, $k =$ antifield number) 
is $\delta$-trivial 
modulo $d$, 
\be 
a_{k}^{p} = \delta \mu_{k+1}^{p} + d \mu_{k}^{p-1} ~ ~ (k \geq 2). 
\label{2.37} 
\ee     
Then, one can always choose $\mu_{k+1}^{p}$ and $\mu_{k}^{p-1}$ to be 
invariant. 
\end{theorem} 
Hence, we have $H^{n,inv}_k(\delta \vert d) = 0$ for $k>2$
while $H^{2,inv}_k(\delta \vert d)$ is given by Theorem \ref{conservation2}.  
 
\subsection{Useful lemmas}
To prove the theorem, we need the 
following three lemmas: 
\begin{lemma} \label{l2.1}  
If $a $ is an invariant polynomial that is $\delta$-exact in the
space of all polynomials, $a = \d b$, 
then, $a $ is also $\delta$-exact in the space of invariant polynomials. 
That is, one can take $b$ to be invariant. 
\end{lemma}  
{\bf{Demonstration of the lemma}} :  
Any function $f([h], [h^*], [C^*], [\x^*])$ can be viewed as a function 
$f(\tilde{h}, [{\cal W}], [h^*], [C^*], [\x^*])$ 
($[{\cal C}]$ instead of ${\cal W}$ for $n=3$). 
Here, the $\tilde{h}$ denote a complete set of (non-invariant) derivatives 
of $h_{\m \n}$ complementary to $[\cw]$, {\it{i.e.}}, 
$\{ \tilde{h}\} = \{h_{\m \n}  , 
\pa_\rho h_{\m \n}, \cdots\}$, where the $\cdots$ denotes combinations of
derivatives independent from $[{\cal W}]$ (or 
$[{\cal C}]$ for $n=3$).  
The $[{\cal W}_{\a \b \g \d}]$'s are not independent 
because of the linearized Bianchi identities, but this does not affect 
the argument. 
An invariant function is just a function that does not involve 
$\tilde{h}$, so one has (if $f$ is invariant), $f = f_{ \vert \tilde{h}=0}$. 
Now, the  differential $\d$ commutes with the operation of 
setting $\tilde{h}$ to zero.  So, if $a = \d b$ and $a$ is invariant, one 
has $ a = a_{ \vert \tilde{h}=0} = (\d b)_{ \vert \tilde{h}=0} 
= \d (b_{ \vert \tilde{h}=0})$, which proves the lemma 
since $b_{ \vert \tilde{h}=0}$ is invariant. 
$\diamond$ 

To explain and establish the second
lemma,  
we first derive a chain of equations with the same structure as (\ref{2.37}) 
\cite{Barnich:1995mt}. 
Acting with $d$ on (\ref{2.37}), we get $d a_{k}^{p} = - 
\d d \m^{p}_{k+1}$.  Using the lemma and the fact that 
$d a_{k}^{p}$ is invariant, we  
can also write $da_{k}^{p}= -\d a_{k+1}^{p+1}$ with $a_{k+1}^{p+1}$ invariant. 
Substituting this in $d a_{k}^{p} = - 
\d d \m^{p}_{k+1}$, we get 
$\d \left[ a_{k+1}^{p+1}-d \m_{k+1}^{p} \right]=0$. As $H(\d)$ is  
trivial in antifield number $>0$, this yields 
\be 
a_{k+1}^{p+1}=\d \m^{p+1}_{k+2}+d\m^{p}_{k+1} 
\ee 
which has the same structure as (\ref{2.37}). We can then repeat 
the same 
operations, until we reach form-degree $n$, 
\be 
a^{n}_{k+n-p}=\d \m^{n}_{k+n-p+1}+ d \m^{n-1}_{k+n-p}. 
\ee 
  
Similarly, one can go down in form-degree. 
Acting with $\d$ on  (\ref{2.37}), one gets  
$\d a^{p}_{k}=-d (\d \m^{p-1}_{k})$. If the antifield  
number $k-1$ of $\d a^{p}_{k}$ is greater than or equal 
to one (i.e., $k>1$), one can rewrite, thanks to Theorem \ref{2.2}, 
$\d a^{p}_{k}=-d a^{p-1}_{k-1}$ where $a^{p-1}_{k-1}$ is invariant. 
(If $k=1$ we cannot go down and the bottom of the chain is (\ref{2.37}) 
with $k=1$, namely $a_1^p=\d\m_2^p+d\m_1^{p-1}$.) 
Consequently $d \left[ a^{p-1}_{k-1}-\d \m^{p-1}_{k} \right]=0$ and,  
as before, we deduce another equation similar to  (\ref{2.37}) : 
\be 
a^{p-1}_{k-1}=\d \m^{p-1}_{k}+d\m^{p-1}_{k-1}. 
\ee 
Applying $\d$ on this equation the descent continues. This descent stops at 
form degree zero or antifield number one, whichever is 
reached first, i.e., 
\bqn 
&{\rm{either}}&~~a^{0}_{k-p}=\d \m^{0}_{k-p+1} 
\nonumber \\ 
&{\rm{or}}&~~a^{p-k+1}_{1}=\d \m^{p-k+1}_{2}+d \m^{p-k}_{1}. 
\eqn 
Putting all these observations together we can write the entire descent as 
\bqn  
a^{n}_{k+n-p}  &=& \d \m^{n}_{k+n-p+1}+d \m^{n-1}_{k+n-p}  
\nonumber \\  
& \vdots &  
\nonumber \\   
a^{p+1}_{k+1}  &=& \d \m^{p+1}_{k+2}+d \m^{p}_{k+1}  
\nonumber \\ 
a^{p}_{k}  &=& \d \m^{p}_{k+1}+d \m^{p-1}_{k}  
\nonumber \\ 
a^{p-1}_{k-1}  &=& \d \m^{p-1}_{k}+d \m^{p-2}_{k-1}  
\nonumber \\ 
& \vdots &  
\nonumber \\ 
{\rm{either}}~~a^{0}_{k-p}&=&\d \m^{0}_{k-p+1} 
\nonumber \\ 
{\rm{or}}~~a^{p-k+1}_{1}&=&\d \m^{p-k+1}_{2}+d \m^{p-k}_{1} 
\label{chain26}
\eqn  
where all the $a^{p \pm i}_{k \pm i}$ are invariants. 

\begin{lemma} \label{l2.1a}
If one of the $\m$'s in
the chain (\ref{chain26}) is invariant, one
can choose all the other $\m$'s in such a way that they
share this property.
\end{lemma}
{\bf{Demonstration}} : 
The proof was given in \cite{Barnich:1995mt} but we
repeat it here for completeness. 
Let us thus assume that $\m^{c-1}_{b}$ is  
invariant. This $\m^{c-1}_{b}$ appears in  
two equations of the descent :  
\bqn 
a^{c}_{b} &=& \d \m^{c}_{b+1}+d \m^{c-1}_{b}, 
\nonumber \\ 
a^{c-1}_{b-1} &=& \d \m^{c-1}_{b}+ d \m^{c-2}_{b-1} 
\eqn 
(if we are at the bottom or at the top, $\m^{c-1}_{b}$ occurs 
in only one equation, and one should just proceed from that one). 
The first equation tells us that $ \d \m^{c}_{b+1}$ is invariant. Thanks 
to Lemma \ref{l2.1} we can choose $\m^{c}_{b+1}$ to be invariant. Looking 
at the second equation, we see that $ d \m^{c-2}_{b-1}$ is invariant 
and by virtue of theorem \ref{2.2}, $\m^{c-2}_{b-1}$ can be chosen 
to be invariant since $b-1 \geq 1$. These two $\m$'s appear each one in two 
different equations of the chain, where we can apply the same reasoning. 
The invariance property propagates then to all the 
$\m$'s.     $\diamond$ 
    
The third lemma is : 
\begin{lemma} \label{l2.2}  
If $a^n_k$ is of antifield number $k>n$, then the $\m$'s in (\ref{2.37}) can 
be taken to be invariant. 
\end{lemma} 
{\bf{Demonstration}} : Indeed, if $k>n$, the last 
equation of the descent is $a^{0}_{k-n}=\d \m^{0}_{k-n+1}$. We can, 
using Lemma \ref{l2.1}, choose $\m^{0}_{k-n+1}$ invariant, and so, 
all the $\m$'s can be chosen to have the same 
property.$\diamond$   
   
\subsection{Demonstration of Theorem \ref{2.6}}  
{}From we have just shown, it is sufficient 
to demonstrate Theorem \ref{2.6}  in form degree $n$ and
in the case where the antifield number 
satisfies $k\leq n$.  
Rewriting the top equation (i.e. (\ref{2.37}) with $p=n$) 
in dual notation, we have 
\be 
a_k=\d b_{k+1}+\pa_{\r}j^{\r}_{k},~ (k\geq 2). 
\label{2.44} 
\ee  
We will work by induction on the antifield number, showing that if  
the property 
is true for antighost numbers $\geq k+1$ (with $k>0$), 
then it is true for $k$.  
As we already know that 
it is 
true in the case $k>n$, the theorem will be demonstrated. 

The idea of the proof is to reconstruct $a_k$ from its
Euler-Lagrange (E.L.) derivatives using the homotopy
formula
\be
a_k=\int^{1}_{0}dt[\frac{\d^La_k}{\d C^{*\a}}(t)C^{*\a}+
\frac{\d^La_k}{\d \x^*}(t)\x^*+
\frac{\d^La_k}{\d h^{*\m\n}}(t)h^{*\m\n}+
\frac{\d^La_k}{\d h_{\m\n}}(t)h_{\m\n}] + \pa_\m V^\m,
\label{homotopy}
\ee     
and to control these E.L. derivatives from (\ref{2.44}).

\subsubsection{Euler-Lagrange derivatives of $a_k$}

Let us take the E.L. derivatives of (\ref{2.44}).
A direct calculation yields
\be 
\frac{\d^La_k}{\d C^{*\a}} =\d Z_{k-1\vert \a}, 
\label{2.45} 
\ee 
with  
$Z_{k-1\vert \a}=\frac{\d^Lb_{k+1}}{\d C^{*\a}}$.
Similarly, 
\be
\frac{\d^La_{k}}{\d \x^*}=\d Z_{k-1},
\label{2.4545}
\ee
with $Z_{k-1}=\frac{\d^Lb_{k+1}}{\d \x^*}$.

For the E.L. 
derivatives with respect to $h^{*\m\n}$ we obtain
\be 
\frac{\d^La_k}{\d h^{*\m\n}}=-\d X_{k\vert\m\n}+2\pa_{(\m}Z_{\n)\vert k-1}
+2\h_{\m\n}Z_{k-1}. 
\label{2.46} 
\ee    
with $X_{k\vert\m\n}=\frac{\d b_{k+1}}{\d h^{*\m\n} }$. 
{}Finally, 
let us compute the E.L. derivatives of $a_k$ with respect to the  
fields.  We get :  
\be 
\frac{\d^La_k}{\d h_{\m\n}}=\d Y^{\m\n}_{k+1}+\cd^{\m\n\r\s}X_{k\vert\r\s},
\label{2.47} 
\ee 
where $Y^{\m\n}_{k+1}=\frac{\d^Lb_{k+1}}{\d h_{\m\n}}$ and where
$\cd^{\a\b\r\s}$ is  the differential operator appearing in
the equations of motion,
\be
\frac{\d\cl_0}{\d h_{\r\s}} = \cd^{\r\s\a\b}h_{\a\b}.
\ee
One has 
\be
\cd^{\m\n\r\s}X_{k\vert\r\s}=\pa_{\a}B^{\a\m\n}_k
\label{5.15}
\ee
for some $B^{\a\m\n}_k$ such that $B^{\a\m\n}_k=-B^{\m\a\n}_k$ and
$\h_{\m\n}B^{\a\m\n}_k=0$.

On the other hand,
since $a_k$ is an invariant object, it depends only on
the linearized Weyl tensor (Cotton tensor for $n=3$) and
its derivatives.  Using the chain rule, one can
thus also rewrite the E.L. derivatives of $a_k$
with respect to the fields as :

\be
\frac{\d^La_k}{\d h_{\m\n}}=\pa_{\a}U_k^{\a\m\n},
\label{ak}
\ee
for some invariant $U_k^{\a\m\n}$ such that $U_k^{\a\m\n}=-U_k^{\m\a\n}$
and $\h_{\m\n}U_k^{\a\m\n}=0$.

Because the E.L. derivatives of $a_k$  are invariant, one can replace,
thanks to lemma \ref{l2.1}, the quantities
$Z_{k-1\vert\a}$, $Z_{k-1}$,  $X_{k\vert \m\n}$ and $Y^{\m\n}_{k+1}$ 
appearing in Eqs. (\ref{2.45}), (\ref{2.4545}), (\ref{2.46}) and 
(\ref{2.47})  
by invariant quantities, i.e.,
rewrite Eqs. (\ref{2.45}), (\ref{2.4545}), (\ref{2.46}) and (\ref{2.47}) 
as
\begin{eqnarray}
\frac{\d^La_k}{\d C^{*\a}}&=& \d Z^{'}_{k-1\vert\a}~, 
\label{2.45bis}
\\
\frac{\d^La_k}{\d \x^{*}}&=&\d Z^{'}_{k-1}~,
\label{2.45ter}
\\
\frac{\d^La_k}{\d h^{*\m\n}}&=&-\d X^{'}_{k\vert\m\n}
+2 \pa_{(\m}Z^{'}_{\n)\vert k-1}+2\h_{\m\n}Z^{'}_{k-1}~,
\label{2.46bis}
\\
\frac{\d^La_k}{\d h_{\m\n}}&=&\d Y^{'\m\n}_{k+1}
+\cd^{\m\n\r\s}X^{'}_{k\vert\r\s}~,
\label{2.47bis}
\end{eqnarray}
for some $Z^{'}_{k-1\vert\a}$, $Z^{'}_{k-1}$, $X^{'}_{k\vert\m\n}$ and 
$Y^{'\m\n}_{k+1}$  that are invariant 
(and not necessarily equal to the
E.L. derivatives
of $b_{k+1}$ any more).
Without loss of generality, one can assume that $Y^{'\m\n}_{k+1}$
is traceless because one has 
$\h_{\m\n}\d Y^{'\m\n}_{k+1}=\d(\h_{\m\n}Y^{'\m\n}_{k+1})=0$
from (\ref{5.15}), (\ref{ak}) and (\ref{2.47bis})
so, by the triviality of $H(\d)$, we can write
$\h_{\m\n}Y^{'\m\n}_{k+1}=\d \psi_{k+1}$
where $\psi_{k+1}$ is invariant by lemma \ref{l2.1}.
Thus one can redefine in (\ref{2.47bis})
$Y^{'\m\n}_{k+1}\rightarrow Y^{'\m\n}_{k+1}-(1/n)\h^{\m\n}\d
\psi_{k+1}$
if necessary. 

\subsubsection{Analysis of $Y^{'\m\n}_{k+1}$}

Now, the tensor $Y^{'\m\n}_{k+1}$ is not only invariant,
but, in view of  
  (\ref{2.47bis}), (\ref{5.15}), (\ref{ak}) and the structure
of $\cd^{\m\n\r\s}$, it fulfills also
\be
\d Y^{'\m\n}_{k+1}=\pa_{\a}M_k^{\a\m\n}
\label{2.50}
\ee
for some invariant $M_k^{\a\m\n}$ such that
$M_k^{\a\m\n}$ is antisymmetric in $\a$, $\m$ and $\h_{\m\n}M_k^{\a\m\n}=0$.
This severely constrains its form.  

To see this, we first note that
the equation (\ref{2.50}) tells us that for fixed $\n$, $Y'^{\m\n}_{k+1}$ 
is a $\delta$-cocycle modulo $d$, in form degree 
$n-1$ and antifield number $k+1$.
$Y'^{\m\n}_{k+1}$ is thus $\delta$-exact 
modulo $d$ ($H^{n-1}_{k+1}(\delta \vert d) 
\simeq H^n_{k+2}(\delta \vert d) \simeq 0$ because $k>0$), 
$Y'^{\m\n}_{k+1}=\d A^{\m\n}_{k+2} +\pa_{\r}T^{\r\m\n}_{k+1}$  where 
$T^{\r\m\n}_{k+1}$ is antisymmetric in  $\r$ and $\m$. By the induction
 hypothesis, which reads $H^{inv}_{s}(\d \vert d) \simeq 0$
for $s>k$ and thus in particular, $H^{inv}_{k+2}(\d \vert d)\simeq 0$,
$A^{\m\n}_{k+2}$ and  $T^{\r\m\n}_{k+1}$  can be 
assumed to be invariant.
Since $Y^{\m\n}_{k+1}$ is symmetric in $\m$ and $\n$, 
we have also 
$\d A^{[\m\n]}_{k+2} +\pa_{\r}T^{\r[\m\n]}_{k+1}=0$.  The 
triviality  
of $H^{n}_{k+2}(\d \vert d)$ and the theorem 8.2 of \cite{{Barnich:1995db}}
implies again 
that $A^{[\m\n]}_{k+2}$ and 
$T^{\r[\m\n]}_{k+1}$ are trivial, in particular, 
$T^{\r[\m\n]}_{k+1}=\d Q^{\r\m\n}_{k+2}+\pa_{\a}S^{\a\r\m\n}_{k+1}$, 
where $S^{\a\r\m\n}_{k+1}$ is antisymmetric 
in ($ \a, \r$)  and in ($\m,\n$), respectively. 
The induction assumption allows us to choose $Q^{\r\m\n}_{k+2}$  and 
$S^{\a\r\m\n}_{k+1}$  to be invariant.  
Writing  
$E^{\m\a\n\b}_{k+1}=-[S^{\m\a\n\b}_{k+1}+S^{\n\b\m\a}_{k+1}]$  
and computing $\pa_{\a\b}E^{\m\a\n\b}_{k+1}$,  we get finally 
that 
\be 
Y'^{\m\n}_{k+1}=\d F^{\m\n}_{k+2}+\pa_{\a\b}E^{\m\a\n\b}_{k+1} 
\label{2.51} 
\ee 
for some invariant $F^{\m\n}_{k+2}=F^{\n\m}_{k+2}$.
By construction, $E^{\m\a\n\b}_{k+1}$ is invariant and has
the symmetries 
$E^{\m\a\n\b}_{k+1} = E^{\n\b\m\a}_{k+1}$, 
$E^{\m\a\n\b}_{k+1} = E^{[\m\a]\n\b}_{k+1}$ and 
$E^{\m\a\n\b}_{k+1} = E^{\m\a [\n\b]}_{k+1}$.

The tracelessness of $Y'^{\m\n}_{k+1}$ implies
$0 = \d F_{k+2}+
\pa_{\a\b} E^{\a \b}_{k+1}$
where we have set
$E^{\a \b}_{k+1} = \eta_{\m \n} E^{\m\a\n\b}_{k+1}$
and
$F_{k+2} = F^{\m}_{k+2 \; \m}$.
This implies in turn
$F_{k+2} = \d K_{k+3} + \pa_\a V^\a_{k+2}$ and
$\pa_{\b} E^{\a \b}_{k+1} = - \d V^\a_{k+2} + \pa_{\b} P^{\a \b}_{k+1}$
for some $K_{k+3}$, $V^a_{k+2}$ and 
$P^{\a \b}_{k+1} = - P^{\b \a}_{k+1}$
that are invariant.
This follows again from our induction hypothesis.
The last condition on $E^{\a \b}_{k+1}$ can be rewritten
$\pa_{\b} (E^{\a \b}_{k+1} - P^{\a \b}_{k+1}) + \d U^\a_{k+2} = 0$
from which one infers, using again $H^{inv}_{k+2}(\delta \vert d)= 0$,
$E^{\a \b}_{k+1} - P^{\a \b}_{k+1} = \pa_\rho Q^{\rho \a \b}_{k+1} + \d
\Phi^{\a \b}_{k+2}$
for some invariants with
$Q^{\rho \a \b}_{k+1} = - Q^{\b \a \rho}_{k+1}$.
Taking the symmetric part of this
equation in $\a$, $\b$
($E^{\a \b}_{k+1}$ is symmetric, $P^{\a \b}_{k+1}$
is antisymmetric) gives
\begin{equation}
E^{\a \b}_{k+1} = \frac{1}{2} \pa_\rho (Q^{\rho \a \b}_{k+1}
+ Q^{\rho \b \a}_{k+1}) + \d \Psi^{\a \b}_{k+2}
\label{MDB}
\end{equation}
with $\Psi^{\a \b}_{k+2}=\Psi^{\b \a}_{k+2}$. 

\subsubsection{Finishing the proof}
We can now complete the argument.  Using the homotopy formula 
(\ref{homotopy})
as well 
as the expressions (\ref{2.45bis}), (\ref{2.45ter}), (\ref{2.46bis}),
(\ref{2.47bis}) for 
these E.L. derivatives, we get 
\be 
a_k=\d \Big( \int^{1}_{0}[Z'_{k-1\vert\a}C^{*\a}+Z'_{k-1}\x^* 
+X'_{k\vert\a\b}h^{*\a\b}+Y'^{\m\n}_{k+1}h_{\m\n}]dt \Big)+\pa_{\r}k^{\r}. 
\ee 
The first three 
terms in the argument of $\d$ are manifestly 
invariant.  To handle
the fourth term, we use (\ref{2.51}). The $\d$-exact term 
disappears ($\d^2=0$). The other one yields, after integrating by
parts twice, a term of the form
$\d [\int_0^1dt E_{k+1}^{\m\a\n\b}{\car}_{\m\a\n\b}]
= \d[\int_0^1dt( E_{k+1}^{\m\a\n\b}{\cw}_{\m\a\n\b}+
4E_{k+1}^{\a\b}{\ck}_{\a\b})]$ (see (\ref{decompRiem})).
The first term under the integral is invariant.
The last step consists in
using (\ref{MDB}) to transform $E_{k+1}^{\a\b}{\cal K}_{\a\b}$
into an acceptable form
\be
E_{k+1}^{\a\b}{\cal K}_{\a\b}\sim - Q^{\rho \b \a}_{k+1}\pa_\rho   
{\cal K}_{\a \b} \sim - Q^{\rho \b \a}_{k+1}\pa_{[\rho}{\cal K}_{\a]\b},
\end{equation}
which shows that this term is also invariant, because the antisymmetrized
derivatives of ${\cal K}_{\a\b}$ are proportional to the linearized
Cotton tensor.

This proves the theorem. 
 
%%%%%%%%%%%%%%%%%%%%%%%%%%%%%%%%%%%%%%%%%%%%%%%%%%%%%%%%%%%%%%%%%%%%%%%%%%%%% 
\section{First-order consistent interactions : $H(s \vert d)$}  
\label{A3} 
%%%%%%%%%%%%%%%%%%%%%%%%%%%%%%%%%%%%%%%%%%%%%%%%%%%%%%%%%%%%%%%%%%%%%%%%%%%%%
\setcounter{equation}{0}  
\setcounter{theorem}{0}  
\setcounter{lemma}{0} 
We have now developed all the necessary tools for the study of the cohomology 
of  $s$ modulo $d$ in form degree $n$. A cocycle of $H^{0,n}(s \vert d)$  
must obey  
\be  
s a + d b = 0 . 
\label{cocycsd}
\ee  
Furthermore, $a$ must be of form degree $n$ and of ghost number $0$.
To analyse (\ref{cocycsd}), we
expand $a$ and $b$ according to the antifield number, 
$a = a_0 + a_1 + ...+a_k  $, 
$b = b_0 + b_1 + ...+b_k  $,
where, as shown in \cite{Barnich:1995mt} and explicitly proved 
in appendix \ref{A2}, 
the expansion stops at some finite antifield number.  
We recall \cite{Henneaux:1997bm} (i) that the antifield-independent piece $a_0$
is the deformation of the Lagrangian; (ii) that $a_1$, which is linear
in the antifields $h^{*\m \n}$, contains the information about the
deformation of the gauge symmetries, given by the
coefficients of $h^{*\m \n}$; (iii) that $a_2$ contains the information
about the deformation of the gauge algebra (the term $C^{*}_A f^A_{\; \; BC}
C^B C^C$ (with $C^{*}_A \equiv C^{* \m}, \x^*$ and $C^A \equiv C_\m, \x$)
gives the deformation of the structure functions appearing
in the commutator of two gauge transformations, while the term
$h^* h^* C C$ gives the on-shell terms); and (iv) that the $a_k$ ($k>2$) give
the information about the deformation of the higher order structure functions, which
appear only when the algebra does not close off-shell.

Writing $s$ as the sum of $\gamma$ and $\delta$, the equation $s a + d b = 0$  
is equivalent to the system of equations $\delta a_i + \gamma a_{i-1}
+ db_{i-1} = 0$ for $i = 1, \cdots, k$, and $\gamma a_k +
db_k =0$.  

\subsection{Deformation of gauge algebra}
Let us assume $k \geq 2$ (the cases $k < 2$ will be
discussed below).  Then, using
the consequence of Theorem \ref{2.2}, one may redefine $a_k$ and $b_k$
so that $b_k = 0$, i.e., $\gamma a_k =0$.   
Then, $a_k = \alpha_J \omega^J$ (up to trivial terms), where the $\alpha_J$
are invariant polynomials and where the $\{\omega^J\}$ form
a basis  of polynomials in $\x, \pa_\m \x, C_\m, \pa_{[\m}C_{\n]}$.  
Acting with  
$\gamma$ on the second to last equation and using $\gamma ^2 =0$~,   
$\gamma a_k = 0$~, we get $d \gamma b_{k-1} = 0$~{\it{i.e.}}
$\g b_{k-1}+dm_{k-1}=0 $
; and then, thanks again
to the  
consequence of theorem  
\ref{2.2}, $b_{k-1}$ can also be assumed to  
be invariant,  
$b_{k-1} = \beta _J \omega ^J$. Substituting these expressions
for $a_k$ and $b_{k-1}$ in the second to last equation, we get :  
\be  
 \delta [\alpha_J \omega^J] + D[\beta_J \omega^J] = \gamma (\ldots).  
\label{2.53}  
 \ee  
As above, this  
equation implies  
\be  
\delta [\alpha_J] \omega^J + D[\beta_J]  \omega^J
\pm \beta_J A^J_I \omega^I = 0
\label{2.54}  
\ee  
since the only combination $\l_J \omega^J$ (with
$\l_J$ invariant) that is $\d$-exact 
vanishes. 
We now expand this equation according to the $D$-degree.  The term  
of degree zero reads  
\be  
[\delta \alpha _{J_0} + D_0 \beta _{J_0}]\omega ^{J_0} =0.  
\ee  
This equation implies that the coefficient of $\omega ^{J_0}$ must be zero,
and  
as $D_0$ acts on the objects upon which $\beta _J$ depends in the same way as
$d$,  
we get :  
\be  
\delta \alpha _{J_0} + d \beta _{J_0} =0.  
\ee  
If the antifield number of $\alpha_{J_0}$ is strictly  
greater than $2$, the solution  
is trivial, thanks to our results on the cohomology  
of $\delta$ modulo $d$:  
\be  
\alpha_{J_0} = \delta \mu_{J_0} + d\nu_{J_0}.  
\ee  
{}Furthermore, theorem \ref{2.6} tells us that   
$\m_{J_0}$ and $\n_{J_0}$ can be chosen  
invariants. We thus get~:  
\bqn  
a_{k}^{0}&=&(\d \m_{J_0} +D_0 \n_{J_0})\o^{J_0}  
\nonumber \\  
&=& s(\m_{J_0}\o^{J_0}) + d (\n_{J_0} \o^{J_0}) + \hbox{"more"}  
\eqn  
where "more" arises from $d \o^{J_0}$ , which can be written as  
$d \o^{J_0} = A^{J_0}_{J_1} \o^{J_1} + su^{J_0}$.  The term   
$\n_{J_0} A^{J_0}_{J_1} \o^{J_1}$ has $D$-degree one, while the  
term $\n_{J_0} s u^{J_0}$ differs from the $s$-exact term  
$s (\pm \n_{J_0}  u^{J_0})$ by  the term $\pm \d(\n_{J_0}) u^{J_0}$,  
which is of lowest antifield number.  
Thus, trivial redefinitions enable one to assume that $a^0_k$ vanishes.  
Once this is done, $\b_{J_0}$ must fulfill $d \b_{J_0} = 0$ and thus  
be $d$-exact in the space of invariant polynomials  
by theorem \ref{2.2} , which allows one to set it to zero  
through appropriate redefinitions.  
  
We can then successively remove the terms of higher $D$-degree  
by a similar procedure, until one has completely redefined away  
$a_k$ and $b_{k-1}$.  One can next repeat the argument for  antifield  
number $k-1$, etc, until one reaches antifield number $2$.
This case deserves more attention, but what we can stress already
now is the 
following : 
{\it{we can assume that the expansion  
of}} $a$  {\it{in}} $sa+db=0$ 
 {\it{stops at antifield number $2$ and takes the form}}  
$a = a_0 +a_1 + a_2$  {\it{with}} $b = b_0 + b_1$.  Note that this results
is independent of any condition on the number of derivatives or
of Lorentz invariance.  These requirements have not been used so far.
The crucial ingredient of the proof is that the
cohomological groups $H^{inv}_k(\delta \vert d)$, which controls the
obstructions to remove $a_k$ from $a$, vanish for $k>2$
as shown in the previous section.

Now let us come to the case $k=2$ and expand $a_2$ in power of the $D$
-degree :
\be
a_2=a^m_2+a^{m+1}_2+\ldots+a^M_2=\sum_{i=m}^{i=M}a^i_2=
\sum_{i=m}^{i=M}P_{I_i}\o^{I_i},
\ee 
and
\be
b_1=\sum_{i=m}^{i=M}\b_{I_i}\o^{I_i}.
\ee
The equation $\d [P_{I_i}\o^{I_i}]+D[\b_{I_i}\o^{I_i}]=-\g a'_1$
implies as before $\d [P_{I_i}\o^{I_i}]+D[\b_{I_i}\o^{I_i}]= 0$.
In $D$-degree minimum we thus get
$[\d P_{I_m}+D_0\b^1_{I_m}]\o^{I_m}= 0$
or, equivalently,
$\d P_{I_m}+d\b^1_{I_m}=0$.
The antighost number is 2 so this equation admits a non-trivial solution
in $H_2(\d\vert d)$.
In $D$-degree zero we have $\o^{I_0}=C_\m C_{\n}$ which cannot be
completed into a Lorentz invariant object by multiplication
with a single $C^{*\a}$, so $m>0$.
We must begin with $a_2^{1}$. 
This time, Lorentz invariance allows 
two non-trivial terms, namely $C^{*\m}C^{\n}C_{[\n,\m]}$ and 
$C^{*\m}C_{\m}\x$. The most general $a_2^{1}$ is therefore
$P_{I_1}\o^{I_1}=C^{*\m}(x_1C_{\m}\x+x_2C^{\n}C_{[\n,\m]})$.
The equation for $a_2^{2} = P_{I_2} \o^{I_{2}}$ reads then
\be
\d P_{I_2}+D_0\b_{I_2}+\b_{I_1} A^{I_1}_{I_2}=0
\label{nodeltamodd}
\ee
where $ A^{I_1}_{I_2}\o^{I_2}=D\o^{I_1}$.
Note that this equation (\ref{nodeltamodd}) does not imply
that $P_{I_2}$ belongs to 
$H_2(\d\vert d)$ anymore.
Now, we have
\be
\b_{I_1} A^{I_1}_{I_2}\o^{I_2}=
2(x_1-x_2)h^{*\m\n}C_{\m}\pa_{\n}\x+2x_2h^*C^{\r}\pa_{\r}\x.
\ee
The coefficient of $C^{\r}\pa_{\r}\x$ is $\d$-exact modulo $D_0$ but the
coefficient of $C_{\m}\pa_{\n}\x$ is not
\footnote{Indeed, one has
$h^{*\m\n}=\tilde{h}^{*\m\n}+\frac{\h_{\m\n}}{n}h^*$ and
$h^{*\m\n}C_{\m}\pa_{\n}\x=\tilde{h}^{*\m\n}C_{\m}\pa_{\n}\x$
$+\frac{1}{n}h^* C^{\r}\pa_{\r}\x$. The term linear in the trace of
$h^{*\m\n}$ can be written as the $\d$ of something, but the term
linear in $\tilde{h}^{*\m\n}$ cannot be equal to
$(\d P_{I_2})\o^{I_2}+(D_0\b^1_{I_2})\o^{I_2}$.}.  
Thus we must impose $x_1 = x_2$.  With this condition,
$P_{I_2}$ reads $ P^2_{I_{2}}=-x_1\x^*$
up to a solution of the homogeneous equation $\d P_{I_2}+D_0\b_{I_2}
=0$.  Lorentz invariance forces one to take this solution to
be zero, since it should contain either $C_{\a}\pa_{\b}\x$ or
$\pa_{[\b} C_{\a]} \x$ contracted with one $C^{*\m}$.
Thus, $a_2^2$ is equal to $a_2^2 = -x_1\x^*C^{\r}\pa_{\r}\x$
and $\b_{I_2}=0$.

The equation for $a_2^3$ is then $\d P_{I_3} + D_0\b_{I_3} = 0$,
which implies as above
$a_2^3=x_3C^{*\m}\x\pa_{\m}\x + 
x_4C^{*\m}\pa^{\n}\x C_{[\n,\m]}$,
where $x_3$ and $x_4$ are arbitrary constants.  These constants
are not constrained by the equation in the next $D$-degree ($n = 4$)
because they yield automatically $\d$-exact (modulo $d$) terms.
Furthermore, Lorentz invariance prevents one from 
adding homogeneous
solution of the form $C^{*\m} \pa_{\a}\x \pa_{\b}\x$
at $D$-degree 4, so the
most general possibility for $a_2$ is
\be
a_2=x_1 \big( C^{*\m} C_{\m}\x
+ C^{*\m} C^{\n}C_{[\n,\m]}
- \x^*C^{\r}\pa_{\r}\x \big)
+ x_3C^{*\m}\x\pa_{\m}\x
+ x_4C^{*\m}\pa^{\n}\x C_{[\n,\m]}.
\ee
The term $a_2$ in the deformation $a$ contains the information
about the deformation of the algebra of the gauge
transformations.  The absence of terms quadratic in the 
antifields $h^{* \m \n}$ indicates that the algebra remains
closed off-shell.  This is not an assumption, it is a consequence
of consistency.

So far, we have used only Lorentz-invariance.  If one imposes
in addition the requirement that the deformed Lagrangian should 
contain no more
derivatives than the original Lagrangian, then one
must set $x_3 = x_4 =0$, since these would lead to terms
with $n+2$ derivatives: in $n$ dimensions,
the free Lagrangian contains $n$ derivatives.  The
count of the derivatives proceed as follows: the antifield $C^{*\m}$ counts
for $n-1$ derivatives, while the ghost $\x$ counts
for one derivative and $C_{\m}$ counts for none  (see
appendix).  We shall
thus set $x_3 = x_4 =0$ from now on and redefine $a_2$ by
adding a
$\g$-exact, to get
\be
a_2 = -x_1
\big (C^{*\m} C^{\n}\pa_{\n}C_{\m} 
+ \x^*C^{\r}\pa_{\r}\x\big) .
\label{a2final}
\ee
>From $a_2$, one can read the deformation of the gauge algebra: in the
deformed theory, the commutator of two transformations parametrized
by the vectors $\x_{1}^{\m}$, $\x_{2}^{\m}$ is no longer zero
but closes according to the diffeomorphism algebra
(as it follows from the term $C^{*\m} C^{\n} \pa_\n C_{\m}$ in
$a_2$), while the commutator of a diffeomorphism with
a Weyl transformation is a Weyl transformation
of parameter $C^{\r}\pa_{\r}\x$ (as it follows from the
second term in $a_2$). 

\subsection{Deformation of gauge transformations}
Having $a_2$, one gets $a_1$ from $\delta a_2 +
\g a_1 + d b_1 = 0$.  This gives 
\be
a_1=-x_1h^{*\m\n}[2h_{\m\n}\x+C^{\r}\pa_{\r}h_{\m\n}+\pa_{\m}C^{\r}h_{\r\n}
 +\pa_{\n}C^{\r}h_{\m\r}]
\label{a1}
\ee
up to a solution of the homogeneous equation $\g a'_1 + db'_1 = 0$.
The analysis of this homogeneous equation is precisely the case $k=1$ mentioned
at the beginning of the previous section.
One can assume $\g a'_1 = 0$ by 
using
the consequence of Theorem \ref{2.2}.  Thus, $a'_1 = \a_J \o^J$,
where the $\a_J$ are invariant polynomials,
$\a_J = \a_J([{\cal W}_{\a \b \m \n}], [h^{*\m\n}])$ and the $\o^J$ are linear
in the ghosts and their (cohomologically non-trivial) derivatives. 
The $\a_J$ must also be linear in the
antifields $h^{*\m\n}$ and their derivatives.  Counting derivatives,
one sees that Lorentz invariance prevents such a term ($h^{*\m\n}$
counts for $n-1$ derivatives and the Weyl tensor for $4$; the only
term with $n$ derivatives that matches the requirements is
$h^{*\m\n} \pa_{[\m} C_{\n]}$, which identically vanishes).

The term $a_1$ yields the 
$\co (\a)$ term
of the full non-linear gauge transformations
(\ref{transfononlin}).

\subsection{Deformation of Lagrangian}
We now restrict the analysis to four spacetime dimensions.
Lifting (\ref{a1}),  we get for $a_0$
the cubic vertex of the Weyl action 
(\ref{actionconf}), 
\bqn
2a_0 &=&-x_1[-4{\cw_{\a\b\g\d}}{\cw_{\a'\b'\g'\d'}}
h^{\a\a'}\h^{\b\b'}\h^{\g\g'}
\h^{\d\d'}+
\nonumber \\
&&+(2\stackrel{(2)}{W}_{\a\b\g\d}
+\frac{h}{2}{\cw_{\a\b\g\d}})
{\cw_{\a'\b'\g'\d'}}\h^{\a\a'}\h^{\b\b'}\h^{\g\g'}\h^{\d\d'}] 
\label{Weylcubic}
\eqn
modulo $\bar{a}_{0}$ solution of the equation
$\g \bar{a}_{0}+db_0=0 $.  
${\cw_{\a\b\g\d}}$ is given in (\ref{linearizedWeylbis}) 
and the conformal Weyl 
tensor at second order is 
\bqn
{\stackrel{(2)}{W}}_{\a\b\g\d}&=&{\stackrel{(2)}{R}}_{\a\b\g\d}
-\frac{2}{n-2}\Big( \h_{\a [\g}{\stackrel{(2)}{R}}_{\d ]\b}+h_{\a [\g}
{\stackrel{(1)}{R}}_{\d ]\b}-\h_{\b [\g}{\stackrel{(2)}{R}}_{\d ]\a}
-h_{\b[\g}{\stackrel{(1)}{R}}_{\d]\a}\Big)
\nonumber \\
&&+\frac{2}{(n-1)(n-2)}\Big( \h_{\a[\g}\h_{\d]\b}\stackrel{(2)}{R} 
+h_{\a[\g}\h_{\d]\b}{\stackrel{(1)}{R}}+\h_{\a[\g}h_{\d]\b}{\stackrel{(1)}{R}}
\big),
\eqn
where ${\stackrel{(2)}{R}}_{\a\b\g\d}$, ${\stackrel{(2)}{R}}_{\m\n}$
and ${\stackrel{(2)}{R}}$ are respectively the Riemann tensor, the Ricci
tensor and the scalar curvature to second order (with all indices down). 
Now, any $\bar{a}_{0}$ solution of $\g \bar{a}_{0}+db_0=0 $
 has Euler-Lagrange
derivatives $\frac{\delta\bar{a}_{0}}{\delta h_{\a \b}}$ which
are (i) invariant; (ii) trace-free; and (iii) divergence-free.
In arbitrary spacetime dimensions, there are many candidates.
However, in $4$ dimensions, there is only one candidate with
$4$ derivatives, namely the Bach tensor.  This corresponds
to a $\bar{a}_{0}$ proportional to the original Lagrangian, 
$\bar{a}_{0} \sim {\cal L}_0$, which can be removed by redefinitions.
Thus, in $4$ dimensions, we can assume $\bar{a}_{0} = 0$, and
there is only one first-order 
consistent deformation that matches all the requirements (Lorentz
invariance, number of derivatives), namely (\ref{Weylcubic}).

Once the first-order vertex has been shown to be unique and has
been identified with the first order Weyl deformation, it
is easy to show that the action can be 
completed to all orders
in the deformation parameter $\alpha$ (we absorb $x_1$ in
$\alpha$
through the redefinition  $-\a x_1\rightarrow \a$).
The argument proceeds as in the case of Einstein
gravity \cite{BDGH} and leads uniquely to the complete Weyl action.

\section{Consistent couplings for different types of Weyl gravitons}
We now turn to the problem of deforming the action (\ref{many})
describing a collection of free ``Weyl fields" $h^a_{\m \n}$.
The computation of the intermediate cohomologies $H(\g)$, $H^{inv}(\d 
\vert d)$ etc proceeds as above, so we immediately go to the calculation
of $H(s \vert d)$ in form degree $n$ and ghost number zero.
Again, one gets that the most general cocycle can be brought
to the form $a = a_0 + a_1 + a_2$ up to trivial terms.
This is because the obstructions to removing $a_k$ ($k>2$) are
absent since $H^{inv}_k(\d
\vert d) = 0$ for $k>2$.

The discussion of the cross-couplings of Weyl gravitons follows very much
the same pattern as the analysis of multi Einstein gravitons \cite{BDGH}.

First we expand $a_2$ as before in power of $D$-degree. The equation 
$\d [P_{I_i}\o^{I_i}]+D[\b_{I_i}\o^{I_i}]=-\g a'_1$
implies  $\d [P_{I_i}\o^{I_i}]+D[\b_{I_i}\o^{I_i}]= 0$.
The most general $a^1_2$ is

\be
P_{I_1}\o^{I_1}=C^{*\m}_{a}(a^a_{bc} C^b_{\m}\x^c+b^a_{bc} C^{b \n}
\pa_{[\m}C^c_{\n]}).
\ee
With this $a^1_2$ we get the following expression for
$\b_{I_1}A^{I_1}_{I_2}\o^{I_2}$ :
\bqn
\b_{I_1}A_{I_2}^{I_1}\o^{I_2}&=&-2  h^*_a \x^b \x^c a^a_{b c} + 
2(a^a_{b c}-b^a_{b c})h^{*\m\n}_a C^b_{\m}\pa_{\n}\x^c +
\nonumber \\
&&+2 h^{*\m}_{a~\r}\pa^{[\r]}C^{\n]b}\pa_{[\m}C_{\n]}^c b^a_{b c}
+2 h^*_a C^{b \n}\pa_{\n}\x^c b^a_{b c}.
\eqn
The equation for $a_2^{2} = P_{I_2} \o^{I_{2}}$ reads 
(see(\ref{nodeltamodd}))
$\d P_{I_2}+D_0\b_{I_2}+\b_{I_1} A^{I_1}_{I_2}=0$. 
For all $I_2$, we must be able to write $\b_{I_1}A_{I_2}^{I_1}$ as a 
$\d$-exact term modulo $D_0$.
Before, we had to impose
$x_1=x_2$ for this equation to have a solution. In this new case with 
$\b_{I_1}A_{I_2}^{I_1}\o^{I_1}$ given above, we have to impose the 
following two conditions
\begin{itemize}
 \item
   $a^a_{bc}=b^a_{bc}$;
 \item
   $a^a_{bc}=a^a_{cb}$.
\end{itemize}
The last condition means that if we view the constants $a^a_{bc}$ 
as defining a product in internal space, then we have a commutative algebra, 
as in  \cite{BDGH}.
When these conditions are fulfilled, we obtain
\be
a_2^2=-\x^*_a C^{b\n}\pa_{\n}\x^c a^a_{bc},
\ee
yielding
\be
a_2=(-\x^*_a C^{b\n}\pa_{\n}\x^c -  C^{*\m}_a  C^{b\n}\pa_{\n}
C^c_{\m}) a^a_{bc}. 
\ee
where we added a $\g$-exact term to simplify the expression
and, as before, assumed that the deformed Lagrangian 
does not possess more derivatives than the original one.

Once $a_2$ is known, one obtains for $a_1$
\be
a_1=-a^a_{\;\; bc}h^{*\m\n}_a[2h^b_{\m\n}\x^c+C^{c\r}\pa_{\r}h^b_{\m\n}
+h^b_{\r\n}\pa_{\m}C^{c\r}+h^b_{\m\r}\pa_{\n}C^{c\r}].
\ee   
It remains then to determine $a_0$.  It turns out that just as
in the Einstein case \cite{BDGH}, there is an obstruction to
the existence of $a_0$, which disappears only if the constants
$a_{abc} \equiv \d_{ad} a^d_{\; bc}$ are completely symmetric.
This means that the commutative algebra defined by the
$a^a_{\; bc}$ should be ``symmetric" \cite{BDGH}.  The metric $\d_{ad}$
that appears here is the metric in internal space defined by
the free (quadratic) Lagrangian (\ref{many}).

To see the appearance of the obstructions, it is enough to focus
on the following two types of terms: (i) terms involving quadratically
the variables of the sector number 1 and linearly the variables
of the sector 2; and (ii) terms involving linearly the variables of the
sectors 1,2 and 3.  Indeed, the other sectors are treated in the same way
(and do not interfere with each others since the numbers $N_i$
counting the variables of the various sectors commute with all the
differentials in the problem).

\begin{enumerate}
\item {\bf Terms of the form ``$(h^1_{\a\b})^2 h_{\g \d}^2$"}.

These terms are determined by two constants, namely
$a^1_{\; 1 2} = a^1_{\; 2 1}$ and $a^2_{\; 1 1}$.
It is easy to verify that the construction is unobstructed if
these constants are equal, $a^1_{\; 1 2} = a^2_{\; 1 1}$.
If there were another choice of these constants that is
unobstructed, then, since the problem is linear, any choice would
be unobstructed.  In particular, $a^2_{\; 1 1} = 0$ would
be acceptable.  However, it is easy to see that 
the choice $a^2_{\; 1 1} = 0$
{\em is} obstructed.  Thus, the only acceptable choice
is the completely symmetric one, $a^1_{\; 1 2} = a^2_{\; 1 1}$.
That $a^2_{\; 1 1} = 0$ leads to an
obstruction follows from a direct
calculation: with $a^2_{\; 1 1} = 0$, $a_1$ reads (in the
$(1)^2$-$2$-sector)
\begin{eqnarray}
a_1 &=& h_1^{*\m\n}[2h^1_{\m\n}\x^2 + 2h^2_{\m\n}\x^1
- C^{1\r}(\pa_{\n}h^2_{\m\r}+\pa_{\m}h^2_{\n\r}
 -\pa_{\r}h^2_{\m\n})  \nonumber \\
& & \; \; \; \; \; \; \; \; \; \; \; \;
\; \; \; \; \; \; \; \; \; \; \; \; -
C^{2\r}(\pa_{\n}h^1_{\m\r}+\pa_{\m}h^1_{\n\r}
 -\pa_{\r}h^1_{\m\n})],
\end{eqnarray}
and yields 
\begin{eqnarray}
\d a_1 &=& {\cal B}_1^{*\m\n}[2h^1_{\m\n}\x^2 + 2h^2_{\m\n}\x^1
- C^{1\r}(\pa_{\n}h^2_{\m\r}+\pa_{\m}h^2_{\n\r}
 -\pa_{\r}h^2_{\m\n})  \nonumber \\
& & \; \; \; \; \; \; \; \; \; \; \; \;
\; \; \; \; \; \; \; \; \; \; \; \; -
C^{2\r}(\pa_{\n}h^1_{\m\r}+\pa_{\m}h^1_{\n\r}
 -\pa_{\r}h^1_{\m\n})].
\end{eqnarray}  
Up to a total derivative, this term must be equal to $\g a_0$.
Without loss of generality, one can assume $a_0 = 
A^{\m \n}([h^1_{\a \b}]) h^2_{\m \n}$ and thus the coefficient
of $h^2_{\m \n}$ in $\g a_0$ is equal to $\g A^{\m \n}$.
But the coefficient of $h^2_{\m \n}$ in $\d a_1$ is (after
integration by parts to remove its derivatives and
up to manifestly $g$-exact terms) equal
to $D {\cal B}_1^{\m\n} \x^1 + {\cal B}_{1 \; \; \r}^\m C_1^{[\n,\r]} 
+ {\cal B}_{1 \; \; \r}^\n C_1^{[\m,\r]} - \pa_\r {\cal B}_1^{\m\n} 
C_1^\r$, which is not $\g$-exact.  Hence, the construction
of $a_0$ is obstructed, as announced.
\item {\bf Terms of the form ``$h_{\a\b}^1 h_{\g  \d}^2 
h_{\l \m}^3$"}

Again, there is a choice of the three constants
$a^1_{\; 2 3} = a^1_{\; 3 2}$, $a^2_{\; 1 3} = a^2_{\; 3 1}$
and $a^3_{\; 1 2} = a^3_{\; 2 1}$ that is unobstructed, namely,
the
completely symmetric one, $a^1_{\; 2 3} = a^2_{\; 1 3} =
a^3_{\; 1 2}$.  If there were a second unobstructed (independent)
choice, then, using linearity, this would imply that there is also
an acceptable choice with, say, $a^3_{\; 1 2} = a^3_{\; 2 1}
=0$.  However, reasoning as above, one easily checks that this is
not the case.  Hence, $a^1_{\; 2 3} = a^2_{\; 1 3} =
a^3_{\; 1 2}$ is the only possibility.
\end{enumerate}

With coupling constants $a_{abc}$ that are completely symmetric,
$a_0$ is given by (\ref{Weylcubic}) with the cubic, 4-derivatives structure
``$\pa\pa\pa\pa h^a h^b h^c a_{abc}$''
\bqn
2a_0 &=& -a_{abc}\Big[ -4{\cw^a_{\a\b\g\d}}{\cw^b_{\a'\b'\g'\d'}}
h^{c\a\a'}\h^{\b\b'}\h^{\g\g'}
\h^{\d\d'}+
\nonumber \\
&&+(2\stackrel{(2)}{W^{ab}}_{\a\b\g\d}+\frac{h^a}{2}{\cw^b_{\a\b\g\d}})
{\cw^c_{\a'\b'\g'\d'}}\h^{\a\a'}\h^{\b\b'}\h^{\g\g'}\h^{\d\d'} \Big].
\label{multiWeylcubic}
\eqn

Proceeding as in \cite{BDGH}, one then finds that there is no
obstruction at second order in the deformation parameter if
and only if the $a^a_{\; bc}$ fulfill the identity
\be
a^a_{\; b[c}a^b_{\; d]f}=0,
\ee
which expresses that the algebra that they define is not
only commutative and symmetric, but also associative.
Since the only such algebras are trivial (direct sums of one-dimensional
ideals) when the internal metric is definite positive, one
concludes that cross-couplings can be removed, exactly as in
\cite{BDGH}.

To restate this result: {\it{there is no possibility of consistent 
cross-couplings 
(with number of derivatives $\leq 4$) between the various Weyl fields
$h^a_{\m\n}$ for the free Lagrangian (\ref{many})}}.  

Now, in the case of Weyl gravity, there does not appear to be
any particularly strong reason for taking the free
Lagrangian to be a sum of free Weyl Lagrangians, as in (\ref{many}).
Any other choice, corresponding to an internal metric $k_{ab}$
that need not be definite positive, would seem to be equally good since
the energy is in any case not bounded from below (or above). 
If one allows non positive definite metrics in internal space, then,
non trivial algebras of the type studied in \cite{CW,W2,Anco}
exist and lead to non trivial cross interactions among the various
types of Weyl gravitons. Rather than developing the general theory,
we shall just give an example with two Weyl fields
$h^1_{\m \n}$, $h^2_{\m \n}$ and metric $k_{ab}$ in internal
space given by
\begin{equation}
\hspace{1.5cm} k_{ab} = \left( \matrix{0 &1 \cr
1 & 0 \cr
} \right). 
\end{equation}    
The complete, interacting action reads
\be
S[h^a_{\m \n}] = \int d^4 x \sqrt{-g} \, h^2_{\m \n} \, B^{\m \n},
\label{strange0}
\ee
where $B^{\m \n}$ is the (complete) Bach tensor of the metric
$g_{\a \b} = \eta_{\a \b} + \a h^1_{\a \b}$ and $g$ its
determinant.
The complete gauge transformations are
\begin{eqnarray}
\frac{1}{\alpha} \d_{\h^a,\f^a}g_{\m\n}
&=&\h^1_{\m;\n}+\h^1_{\n;\m}+2\f^1 g_{\m\n} ,
\label{strange1}\\
\d_{\h^a,\f^a} h^2_{\m \n} &=& \a {\cal L}_{\h^1} h^2_{\m \n}
+ 2 \a \f^1 h^2_{\m \n} + \h^2_{\m;\n}+\h^2_{\n;\m}
+2\f^2 g_{\m\n}, 
\label{strange2}
\end{eqnarray}
where covariant derivatives ($;$) are computed with the 
metric $g_{\a \b}$.  The invariance of the
action (\ref{strange0}) under (\ref{strange1}) and
(\ref{strange2}) is a direct consequence of the
identities fulfilled by the Bach tensor, i.e.
\be
B^{\a \b}_{\; \; ;\b} = 0, \; B^{\a \b} g_{\a \b} = 0, \;
\d_{\h^a,\f^a} B^{\a \b} = {\cal L}_{\a\h^1} B^{\a \b} -
6 \a \f^1 B^{\a \b}.
\ee

The theory with action (\ref{strange0}) can probably be
given a nice group-theoretical interpretation along the
lines of \cite{KTvN1}, which we plan to investigate.
This would be useful for its supersymmetrization.
[A complete action and transformation rules
for conformal supergravity was given in \cite{KTvN2}.
Some simplifications of the results were presented
in \cite{TvN}.  The simplest treatment is given
in \cite{PvN}. A nice feature of $N=1$ conformal
supergravity is that the gauge algebra closes
without auxiliary fields.]

%%%%%%%%%%%%%%%%%%%%%
\section{Conclusions}
%%%%%%%%%%%%%%%%%%%%%
%
\setcounter{equation}{0}  
\setcounter{theorem}{0}  
\setcounter{lemma}{0} 
In this paper, we have found results about the uniqueness of Weyl gravity
in four dimensions.
Our method relies on the antifield approach and uses cohomological 
techniques.

\subsection{Deformations of the gauge transformations}
We have found all the possible deformations of the gauge transformations
of linearized
Weyl gravity, under the assumptions that
the deformed Lagrangian is Lorentz invariant and contains no more 
derivatives than those appearing in the free
Lagrangian.  We have shown that 
the only possibility is given by 
\be
\frac{1}{\alpha} \d_{\h,\f}g_{\m\n}=\h_{\m;\n}+\h_{\n;\m}+2\f g_{\m\n},
~~~~~g_{\m\n}=\h_{\m\n}+\a h_{\m\n}
\label{71.71}
\ee   
(the limit $\a = 0$ corresponds to no deformation at all).
 
If one does not impose the requirement on the number
of derivatives, there are two other possibilities for the deformations
 of the gauge algebra, but we have not
investigated them, although we suspect that they will
be ultimately inconsistent (if only because they
involve coupling constants with negative mass dimensions).

\subsection{Deformations of the  Lagrangian}
We have then derived the most general deformation of the 
Lagrangian, which was shown to 
be just the non-linear Weyl gravity Lagrangian (\ref{actionconf})
in four dimensions.
 
Our results can easily be extended to $3$, or higher, dimensions, because it is only in the last step, i.e., in the
construction of $a_0$, that we used the assumption $n=4$. 

\begin{enumerate}
\item $n=3$: in $3$ spacetime dimensions, the free action
is (\cite{DJT})
\be
S_0^{CS}=\frac{1}{2}\int d^3 x \varepsilon_{\m\a\b}{\stackrel{(1)}{G^{\a\n}}}
 \pa^{\m}h^{\b}_{~\n}
\label{Ln=3}
\ee
 and the free Lagrangian is gauge-invariant only
up to a total derivative.  It contains three derivatives.  Deforming
this theory leads uniquely to the theory of \cite{DJT,HW}.
\item $n = 2p \geq 6$: in even spacetime dimensions, the
free Lagrangian is given by the quadratic part of the unique
conformal invariant that has a non-vanishing quadratic part,
which reads,
(see for example proposition 3.4. p. 106 of \cite{GF})
\be
I_n\propto {\cw}_{\a\b\g\d} {\tilde{\bigtriangleup}}_n^{\a\b\g\d\m\n\r\s}
{\cw}_{\m\n\r\s},
\label{XXYYZZ}
\ee
where $ {\tilde{\bigtriangleup}}_n^{\a\b\g\d\m\n\r\s}$ is a differential
operator
of order $n-4$ in dimension $n$.
There are, however, other conformal invariants that have no quadratic parts. 
These can come into the deformation process, through $a_0$. Therefore, 
deforming the free theory based on (\ref{XXYYZZ}) 
gives not only the full conformally
invariant completion of (\ref{XXYYZZ}), but
also all the other possible conformal
invariants, the number of which grows with the dimension 
(see for instance \cite{GF, PR}). Thus the deformed Lagrangian 
is no longer unique. The other results on the deformation
of the gauge algebra and the gauge transformations are
otherwise unchanged.  
For instance, in dimension $n=6$ we have the invariant at the linearized level
\be
\ci=\pa^{\a}\cw^{\b\g\d\e}\pa_{\a}\cw_{\b\g\d\e}
+ 16 \cc^{\g\d\e}\cc_{\g\d\e}-16\cw^{\a\g\d\e}\pa_{\a}\cc_{\g\d\e}~,
\ee
which, deformed, would give the invariant 
\bqn
\sqrt{-g}I_6&=&\sqrt{-g}[\bigtriangledown^{\a}W^{\b\g\d\e}
\bigtriangledown_{\a}W_{\b\g\d\e}
+ 16 C^{\g\d\e}C_{\g\d\e}-16W^{\a\g\d\e}\bigtriangledown_{\a}C_{\g\d\e}
\nonumber \\
&&+16K_{\a\b}W^{\a\g\d\e}W^{\b}_{~\g\d\e}].
\eqn
The 
other two known terms 
$\Omega_1=\sqrt{-g}W^{\m\s}_{~~\r\n}W^{\r\n}_{~~\a\b}W^{\a\b}_{~~\m\s}$ 
\\
and
$\Omega_2=\sqrt{-g}W^{\m\s\r\n}W_{\n\b\s\a}W^{\a~\b}_{~\r~\m}$
can also come in
through $a_0$.
                                                                   
Furthermore,  there exists the 
possibility of taking $a_2$ (and then also $a_1$) equal to zero,
i.e., of not deforming the gauge transformations at all, while taking
a non-trivial $a_0$ polynomial in the linearized Weyl curvatures
with the appropriate number of derivatives.  In six dimensions,
$\cw^{\m\s}_{~~\r\n}\cw^{\r\n}_{~~\a\b}\cw^{\a\b}_{~~\m\s}$ and
$ \cw^{\m\s\r\n}\cw_{\n\b\s\a}\cw^{\a~\b}_{~\r~\m}$
 would be possible interactions
that do not deform the gauge symmetry.  This possibility does
not exist in four dimensions because the candidate interactions
would contain more than four derivatives.
\end{enumerate}

\subsection{Interactions for a collection of Weyl fields}
We have then investigated interactions for a collection
of Weyl fields and have shown that cross interactions were
impossible in four dimensions with the prescribed
free field limit (\ref{many}), although non
trivial possibilities exist with a different free Lagrangian.
Many of our considerations hold in higher dimensions.
However, one can then also build non trivial cross interactions
that do not modify the gauge structure, e.g., by adding
$g_{abc} {\cal W}^a_{\a \b \g \d} 
{\cal W}_{\; \l \m}^{b \; \; \; \a \b}
{\cal W}^{c \g \d \l \m}$ to the free Lagrangian (in 6 dimensions).

\section*{Acknowledgements}  
We are very grateful to Peter van Nieuwenhuizen for 
fruitful discussions
that initiated this work.
N.B. thanks  Glenn Barnich, Xavier Bekaert and Christiane
Schomblond for discussions and advice.    
This work is partially supported by the ``Actions de  
Recherche Concert{\'e}es" of the ``Direction de la Recherche  
Scientifique - Communaut{\'e} Fran{\c c}aise de Belgique", by  
IISN - Belgium (convention 4.4505.86),  by  
Proyectos FONDECYT 1970151 and 7960001 (Chile)
and by the European Commission RTN programme
HPRN-CT-00131
in which the authors are associated to K. U. Leuven.  

\appendix
%%%%%%%%%%%%%%%%%%
\section{Counting derivatives}
\label{A2}
%%%%%%%%%%%%%%%%%%
\setcounter{equation}{0}  
\setcounter{theorem}{0}  
\setcounter{lemma}{0} 
In this appendix we establish the following result: let $a_0$
be a consistent first-order
deformation of the free Lagrangian with bounded number
of derivatives, i.e., be a solution of
$\g a_0 + \pa_{\m}t^{\m}\approx 0$, or, what is the same
\be
\g a_0+\d a_1 +\pa_{\m}t^{\m}=0
\ee  
for some $a_1$. We assume that the spacetime dimension is such that
$n \geq 3$.
Then the corresponding BRST cocycle $a = a_1+a_2+\ldots$
obtained by completing $a_0$ in such a way that $sa+db=0$
can be assumed to have a finite expansion.

To prove this, one assigns a new degree to the fields,  called
the $K$-degree, 

\be
\begin{tabular}{|c||l|l|l|l|l|l|l|l|l|}
\hline
fields &$h^{*\m\n}$ &$C^{*\m}$ &$\x^*$ &$\pa_{\m}$ &$\d$ &$\g$ 
&$h_{\m\n}$ &$C_{\m}$ &$\x$     \\
\hline\hline
$K$-degree &n-1 &n-1 &n-2 &1 &1 &1 &0 &0 &1    \\
\hline
\end{tabular}
\ee

The $K$-degree counts in fact the dimension (there is some freedom
in the assignments for the ghosts and the antifields; we choose to assign
dimension $1$ to $\g$ and $\d$ for convenience).  Note in particular that
the $K$-degree of an expression involving only $h_{\m \n}$ and its
derivatives is precisely equal to the number of derivatives.

The $K$-degree is increased by one by $\d$,  $\g$,  $\pa_{\m}$, 
so the K-degree is the same for $a_0$, $a_1$, etc.
Because $a_0$ has a bounded derivative order, and one
may assume it to be homogeneous, so $K(a_0)=N$ for 
some finite $N$.
The claim is : $a=a_0+\ldots+a_k$ stops at antifield number $k\leq N$.

Proof : From the above, $K(a_k)=N$.
One has $k=n_{h^*}+2n_{C^*}+2n_{\x^*}$, where $n_{h^*}$ 
(resp. $n_{C^*}$ and $n_{\x^*}$ ) counts the number of $h^{*\m\n}$ 
(resp. $C^{*\m}$ and $\x^*$), differentiated or not.
On the other hand we know that
$K(a_k)\geq (n-1)n_{h^*}+(n-1)n_{C^*}+(n-2)n_{\x^*}$. 
The inequality holds because some differentiations may appear. 
This expression is clearly greater than or equal
to the expression $n_{h^*}+2n_{C^*}+2n_{\x^*}$ for $n \geq 4$
so in these cases we proved that  
\be
k\leq N.
\ee  
In the case $n=3$ 
one has $k\leq 2K(a_k)$ and thus $k\leq 2N$.


\begin{thebibliography}{99}   
\bibitem{BRS} C.~Becchi, A.~Rouet and R.~Stora,
%``Renormalization Of The Abelian Higgs-Kibble Model,''
Commun.\ Math.\ Phys.\  {\bf 42}, 127 (1975).
%%CITATION = CMPHA,42,127;%%
\bibitem{Tyutin} I.~V.~Tyutin,
%``Gauge Invariance In Field Theory 
% And Statistical Physics In Operator Formalism,''
preprint LEBEDEV-75-39, unpublished. 
\bibitem{Barnich:1995db}
G.~Barnich, F.~Brandt and M.~Henneaux,
%``Local BRST cohomology in the antifield formalism. I. General theorems,''
Commun.\ Math.\ Phys.\ {\bf 174} (1995) [hep-th/9405109].
%%CITATION = HEP-TH 9405194;%%
\bibitem{Barnich:1995mt}
G.~Barnich, F.~Brandt and M.~Henneaux,
%``Local BRST cohomology in the antifield formalism. II. Application to
%Yang-Mills theory,''
Commun.\ Math.\ Phys.\ {\bf 174} (1995) 93 [hep-th/9405194].              
%%CITATION = HEP-TH 9405194;%%
\bibitem{BBHEinst} G.~Barnich, F.~Brandt and M.~Henneaux,
%``Local BRST cohomology in Einstein Yang-Mills theory,''
Nucl.\ Phys.\ B {\bf 455}, 357 (1995) [hep-th/9505173].
%%CITATION = HEP-TH 9505173;%% 
\bibitem{Barn-Henn} G.~Barnich and M.~Henneaux,
%``Consistent couplings between fields with a gauge
%freedom and deformations of the master equation,''
Phys.\ Lett.\  {\bf B311}, 123 (1993)
[hep-th/9304057].
%%CITATION = HEP-TH 9304057;%%        
\bibitem{MTW} C.W. Misner, K.S. Thorne and J.A.  Wheeler,
``Gravitation", Freeman (San Francisco: 1973) (chapter 17).
\bibitem{Fierz:1939ix}  
W.~Pauli and M.~Fierz, 
Helv. Phys. Acta {\bf 12}, 297 (1939).
%%CITATION = HPACA,12,297;%%
\bibitem{all} S. N. Gupta,
%{\it Gravitation and Electromagnetism},
Phys. Rev. {\bf 96} (1954) 1683;
%%CITATION = PHRVA,96,1683;%%
R. H. Kraichnan,
%{\it Special relativistic derivation of generally covariant
%gravitation theory},
Phys. Rev. {\bf 98} (1955) 1118;
%%CITATION = PHRVA,98,1118;%%
R.~P.~Feynman, F.~B.~Morinigo, W.~G.~Wagner and B.~.~Hatfield,
``Feynman lectures on gravitation,''
{\it  Reading, USA:
Addison-Wesley (1995) 232 p. (The advanced book program)};
S. Weinberg, Phys. Rev. {\bf 138}, B988 (1965);
V. I. Ogievetsky and I. V. Polubarinov,
%{\it Interacting Field
%of Spin 2 and the Einstein Equations},
Ann. Phys. {\bf 35} (1965) 167;
%%CITATION = APNYA,35,167;%%
W. Wyss, Helv. Phys. Acta {\bf 38}, 469 (1965);
S.~Deser,
%``Self-interaction And Gauge Invariance,''
Gen.\ Rel.\ Grav.\  {\bf 1}, 9 (1970);
%%CITATION = GRGVA,1,9;%%
D. G. Boulware and S. Deser, Ann. Phys. (NY) {\bf 89}, 193 (1975);
%%CITATION = APNYA,89,193;%%
J. Fang and C. Fronsdal, J. Math. Phys. {\bf20}, 2264 (1979);
%%CITATION = JMAPA,20,2264;%%
F.~A.~Berends, G.~J.~Burgers and H.~Van Dam,
%``On Spin Three Self-interactions,''
Z.\ Phys.\  {\bf C24}, 247 (1984);
%%CITATION = ZEPYA,C24,247;%%
R.~M.~Wald,
%``Spin 2 Fields And General Covariance,''
Phys.\ Rev.\  {\bf D33}, 3613 (1986).
%%CITATION = PHRVA,D33,3613;%%
't Hooft and M.~Veltman,
%``One Loop Divergencies In The Theory Of Gravitation,''
Annales Inst. H. Poincar\'e Phys.\ Theor.\ A {\bf 20} (1974) 69.
%%CITATION = AHPAA,A20,69;%%
\bibitem{BDGH}
N.~Boulanger, T.~Damour, L.~Gualtieri and M. Henneaux,
Nucl.\ Phys.\ B {\bf 597}, 127 (2001)[hep-th/0007220];  
%%CITATION = HEP-TH 0007220;%%
{\it{proceedings of International Conference on Quantization, Gauge Theory,
 and Strings: Conference Dedicated to the Memory of Professor Efim
Fradkin, Moscow, Russia, 5-10 Jun 2000}}, [hep-th/0009109]. 
%"No consistent cross-interactions for a collection ofmassless spin-2  fields"
%%CITATION = HEP-TH 0009109;%%"
\bibitem{0007211}
V. Balasubramanian, E. Gimon, D. Minic, J. Rahmfeld,
%``Four Dimensional Conformal Supergravity From AdS Space'',
 [hep-th/0007211].
%%CITATION = HEP-TH 0007211;%%
\bibitem{Schmidt}
H.~J.~Schmidt,
[gr-qc/0105108].
%Non-trivial Solutions of the Bach Equation Exist",
%%CITATION = GR-QC 0105108;%%"
\bibitem{FT}
E. S. Fradkin and A. A. Tseytlin,
     Phys.\ Rept.\ {\bf{119}} 233 (1985).
%%CITATION = PRPLC,119,233;%%
\bibitem{ZinnJust}
J. Zinn-Justin, "Quantum Field Theory And Critical Phenomena",
{\it{Oxford, UK: Clarendon (1989) International series
of monographs on physics, 77.}} Third edition 1996;
J.~Zinn-Justin,
``Renormalization Of Gauge Theories,''
SACLAY-D.PH-T-74-88
{\it Lectures given at Int. Summer Inst.
for Theoretical Physics, Jul 29 - Aug 9, 1974, Bonn, West Germany}. 
\bibitem{dWvH} B.~de Wit and J.~W.~van Holten,
%``Covariant Quantization Of Gauge Theories With Open Gauge Algebra,''
Phys.\ Lett.\ B {\bf 79}, 389 (1978).
%%CITATION = PHLTA,B79,389;%% 
\bibitem{BV}
I. A. Batalin and G. A. Vilkovisky, Phys.\ Rev.\ {\bf{D 28}} (1983) 2567.
%%CITATION = ERRAT,D30,508;%%
\bibitem{M1990}
M. Henneaux, Nucl.\ Phys.\ B (Proc. Suppl.){\bf{18 A}}(1990) 47.
\bibitem{GPS1995}
J. Gomis, J. Paris, S. Samuel, Phys.\ Rep.\ {\bf{259}} (1995) 1.
%%CITATION = HEP-TH 9412228;%%
\bibitem{BOOK} M. Henneaux and C. Teitelboim, {\it Quantization
of Gauge Systems}, Princeton University Press, Princeton: 1992,
\bibitem{Henneaux:1997bm}
M.~Henneaux,
%``Consistent interactions between gauge fields: The cohomological  approach,''
{\it Proceedings of a Conference 
on Secondary Calculus and Cohomological Physics, Aug 24-31, 1997,
Moscow, Russia,}
Cont.\ Math.\ {\bf 219} (1998) 93 [hep-th/9712226].
%%CITATION = HEP-TH 9712226;%%
\bibitem{Stasheff:1997fe}
J.~Stasheff,
%``Deformation theory and the Batalin-Vilkovisky master equation,''
[q-alg/9702012].
%%CITATION = Q-ALG 9702012;%% 
\bibitem{review}
G. Barnich, F. Brandt and M. Henneaux, 
%``Local BRST cohomology in gauge theories,''
Phys.\ Rept.\ {\bf 338}, 439 (2000)
[hep-th/0002245].
%%CITATION = HEP-TH 0002245;%%
\bibitem{S-T} A.~Schwimmer and S.~Theisen,
%``Diffeomorphisms, anomalies and the Fefferman-Graham ambiguity,''
JHEP {\bf 0008}, 032 (2000)
[hep-th/0008082].
%%CITATION = HEP-TH 0008082;%% 
\bibitem{BVPT1}
F. Brandt, W. Troost, A. Van Proeyen, 
%``Background Charges and Consistent
%Continuous Deformations of 2-d Gravity Theories'' 
Phys.\ Lett.\ {\bf{B374}} 31-36,1996  [hep-th/9510195].
%%CITATION = HEP-TH 9510195;%%
\bibitem{BVPT2}
F. Brandt, W. Troost, A. Van Proeyen, 
%``The BRST-Antibracket cohomology of
%2-d gravity conformally coupled to scalar matter'',
Nucl.\ Phys.\ {\bf{B464}} 353-408, 1996, [hep-th/9509035]. 
%%CITATION = HEP-TH 9509035;%%
\bibitem{bach}
R. Bach, 
%``Zur Weylschen Relativit\"atstheorie und der Weylschen Erweiterung
%des Kr\"ummungstensorbegriffs, 
Math.\ Z.\ , t.\ {\bf{9}}, 1921, p. 110-135.  
\bibitem{Duboisetal}
M. Dubois-Violette, M. Henneaux, M. Talon and C. Viallet,
Phys. Lett. {\bf{B267}}, 81 (1991).
%%CITATION = PHLTA,B267,81;%%
\bibitem{FH}
J M Fisch and M. Henneaux, Comm. Math Phys. {\bf{128}}, 627 (1990). 
%%CITATION = CMPHA,128,627;%%
\bibitem{locality}
M. Henneaux, Commun. Math. Phys. {\bf{140}}, 1 (1991).
%%CITATION = CMPHA,140,1;%%
\bibitem{Brandt}
F.~Brandt, 
%''Gauge theories of spacetime interactions''
[hep-th/0105010].
%%CITATION = HEP-TH 0105010;%%"
\bibitem{CW} C. Cutler and R. Wald,
%{\it A new type of gauge invariance for
%a collection of massless spin--$2$ fields/ I. Existence and uniqueness},
Class. Quant. Grav. {\bf 4} (1987) 1267.
%%CITATION = CQGRD,4,1267;%%   
\bibitem{W2} R. Wald,
%{\it A new type of gauge invariance for
%a collection of massless spin--$2$ fields/ II. Geometrical interpretation},
Class. Quant. Grav. {\bf 4} (1987) 1279.   
%%CITATION = CQGRD,4,1279;%%
\bibitem{Anco} S.~C.~Anco,
%``Nonlinear gauge
%theories of a spin-two field and a spin-three-halves  field,''
Annals Phys.\  {\bf 270}, 52 (1998).
%%CITATION = APNYA,270,52;%%       
\bibitem{KTvN1} 
M.~Kaku, P.~K.~Townsend and P.~Van Nieuwenhuizen,
%``Gauge Theory Of The Conformal And Superconformal Group,''
Phys.\ Lett.\ B {\bf 69}, 304 (1977).
%%CITATION = PHLTA,B69,304;%%
\bibitem{KTvN2} M.~Kaku, P.~K.~Townsend and P.~van Nieuwenhuizen,
%``Properties Of Conformal Supergravity,''
Phys.\ Rev.\ D {\bf 17}, 3179 (1978).
%%CITATION = PHRVA,D17,3179;%%
\bibitem{TvN} P.~K.~Townsend and P.~van Nieuwenhuizen,
%``Simplifications Of Conformal Supergravity,''
Phys.\ Rev.\ D {\bf 19}, 3166 (1979).
%%CITATION = PHRVA,D19,3166;%%
\bibitem{PvN} P. van Nieuwenhuizen,
``Gauging of spacetime algebras'', in ``Quantum Groups and their applications
in Physics'',{\it{ Proceedings of the 127th Enrico Fermi School at Varenna, 
June 1996, p. 595}}.
\bibitem{DJT}
S.~Deser, R.~Jackiw and S.~Templeton,
%``Topologically massive gauge theories,''
Annals Phys.\ {\bf 140}, 372 (1982)
[Erratum-ibid.\ {\bf 185}, 372 (1982)].
%%CITATION = APNYA,140,372;%%
\bibitem{HW}
H. Horne, E. Witten, 
%``Conformal Gravity in Three-Dimensions as a Gauge Theory'' 
Phys.\ Rev.\ Lett.\ {\bf{62}} 501-504,1989.
%%CITATION = PRLTA,62,501;%%
\bibitem{GF}
C. Fefferman and C. Graham, Ast\'erisque, hors s\'erie (1985) 95.
\bibitem{PR}
T. Parker and S. Rosenberg, J.\ Diff.\ Geometry\ {\bf{25}} (1987) 199.

\end{thebibliography}
\end{document}